\documentclass[12pt]{amsart}
\usepackage[margin=1.25in]{geometry}
\usepackage{amsmath,amsthm,amsfonts,amssymb}
\usepackage[nosort]{cite}

\newlength{\xtrawidth}
\newlength{\xtraheight}
\numberwithin{equation}{section}
\numberwithin{table}{section}
\numberwithin{figure}{section}
\begin{document}

\title{SU(N) transitions in M-theory on Calabi--Yau fourfolds and background fluxes}

\date{February 25, 2016.  Revised March 23, 2016. Revised February 22, 2017.}

\author{Hans Jockers}
\address{Bethe Center for Theoretical Physics, Rheinische Friedrich-Wilhelms-Universit\"at Bonn, Nussallee 12, D--53115 Bonn, Germany}
\email{jockers@uni-bonn.de}

\author{Sheldon Katz}
\address{Department of Mathematics, University of Illinois at Urbana-Champaign, 1409 W. Green St., Urbana, IL 6180, USA}
\email{katz@math.uiuc.edu}

\author{David R. Morrison}
\address{Departments of Mathematics and Physics, University of California at Santa Barbara, Santa Barbara, CA 93106, USA}
\email{drm@physics.ucsb.edu}

\author{M. Ronen Plesser}
\address{Erna and Jakob Michael Visiting Professor, Weizmann Institute
of Science, Rehovot 76100, Israel;
IHES, Le Bois-Marie 35, route de Chartres 91440
Bures-sur-Yvette, France}
\email{plesser@cgtp.duke.edu}

\begin{abstract}
We study M-theory on a Calabi--Yau fourfold with a smooth surface~$S$ of $A_{N-1}$ singularities. The resulting three-dimensional theory has a $\mathcal{N}=2$ $SU(N)$ gauge theory sector, which we obtain from a twisted dimensional reduction of a seven-dimensional $\mathcal{N}=1$ $SU(N)$ gauge theory on the surface $S$. A variant of the Vafa--Witten equations governs the moduli space of the gauge theory, which --- for a trivial $SU(N)$~principal bundle over $S$ --- admits a Coulomb and a Higgs branch. In M-theory these two gauge theory branches arise from a resolution and a deformation to smooth Calabi--Yau fourfolds, respectively. We find that the deformed Calabi--Yau fourfold associated to the Higgs branch requires for consistency a non-trivial four-form background flux in M-theory. The flat directions of the flux-induced superpotential are in agreement with the gauge theory prediction for the moduli space of the Higgs branch. We illustrate our findings with explicit examples that realize the Coulomb and Higgs phase transition in Calabi--Yau fourfolds embedded in weighted projective spaces. We generalize and enlarge this class of examples to Calabi--Yau fourfolds embedded in toric varieties with an $A_{N-1}$ singularity in codimension two.
\end{abstract}

\maketitle

\section*{Introduction}
The construction of gauge theories via dimensional reduction on Calabi--Yau varieties with singularities has become a powerful approach to study both supersymmetric gauge theories and moduli spaces of Calabi--Yau varieties in the vicinity of singularities. For gauge theories with eight supercharges --- such as $\mathcal{N}=2$ theories in four spacetime dimensions or $\mathcal{N}=1$  theories in five spacetime dimensions --- the interplay between Higgs and Coulomb branches of the gauge theory and the corresponding ``extremal transitions'' between geometric phases of singular Calabi--Yau threefolds has led to important insights into strongly coupled supersymmetric gauge theories and their moduli spaces \cite{Strominger:1995cz,Greene:1995hu,Katz:1996ht,Klemm:1996kv,Witten:1996qb,Morrison:1996xf,Katz:1996fh,Intriligator:1997pq}.  

While holomorphy strongly constrains supersymmetric theories with eight supercharges \cite{deWit:1984pk,Cremmer:1984hj,Strominger:1990pd,Seiberg:1994rs,Seiberg:1994aj}, it is a less powerful tool for supersymmetric theories with four supercharges \cite{Cremmer:1982en,Intriligator:1994jr,Seiberg:1994bp}, e.g., $\mathcal{N}=1$ theories in four spacetime dimensions and $\mathcal{N}=2$ theories in three spacetime dimensions. As a consequence the analysis of the gauge theory branches becomes more challenging but also richer. (For a proposal of a transition in a system with only two supercharges see ref.~\cite{Gukov:2002zg}.)

In this work gauge theories with four supercharges are constructed from M-theory on a Calabi--Yau fourfold. That is to say we want to make predictions regarding the relevant moduli spaces  (and transitions among them) of the low-energy physics governing degrees of freedom localized near a complex codimension two singularity, which gives rise to a three-dimensional $\mathcal{N}=2$ gauge theory, studied for instance in refs.~\cite{Affleck:1982as,Aharony:1997bx,deBoer:1997kr,Diaconescu:1998ua,Kapustin:1999ha,Dorey:1999rb,Tong:2000ky,Aganagic:2001uw,Intriligator:2012ue,Intriligator:2013lca}. This should be contrasted with results obtained from codimension two singularities in type~II string theories for theories with eight supercharges studied in refs.~\cite{Katz:1996ht,Klemm:1996kv}. In both scenarios the essential idea is that such codimension two singularities are associated to non-Abelian enhanced gauge symmetry. This is understood from the duality between M-theory compactified on K3 to the heterotic string compactified on $T^3$ \cite{Witten:1995ex}. The charged degrees of freedom represent M2-branes wrapping the two-cycles whose volume vanishes in the singular limit. In the limit in which the K3 volume is large, bulk modes decouple from the low-energy dynamics of the seven-dimensional modes localized at the singular locus. The resulting gauge theory is described by a non-Abelian gauge theory with sixteen supercharges. Compactifying further on a circle shows that IIA~theory near an ADE singularity exhibits enhanced gauge symmetry in six dimensions. Now the charged modes are associated to wrapped D2-branes.

In a Calabi--Yau $n$-fold --- $n=4$ for the M-theory compactifications to three dimensions or $n=3$ for the IIA compactification to four dimensions ---  a codimension two singularity can be thought of as a face of suitable codimension in the K\" ahler cone, in which some number of divisors shrink to a $(n-2)$-dimensional locus $S$ along which we find an ADE singularity. Deep in the interior of this face, the volume of $S$ and those of any relevant submanifolds are large, which means we can study the low-energy dynamics by a suitably twisted dimensional reduction along $S$ of the gauge theory from seven (respectively six) dimensions down to three (respectively four) dimensions. This will lead to a prediction for an ${\mathcal N}=2$ gauge theory describing the low-energy dynamics near the singular locus.  K\"ahler deformations away from the singularity will describe the Coulomb branch of this theory, while its Higgs branch, when present, will lead to a prediction for the complex structure moduli space of the related compactification on the Calabi--Yau space obtained via an extremal transition. 

Compared to gauge theories with eight supercharges from singular Calabi--Yau threefolds \cite{Katz:1996ht,Klemm:1996kv}, constructing theories with four supercharges from Calabi--Yau fourfolds requires additional geometric data \cite{Becker:1996gj,Sethi:1996es}. Namely, in order to entirely describe the geometrically engineered gauge theory, it is necessary to specify the M-theory compactification on the Calabi--Yau fourfold together with a suitable four-form background flux. That is to say, the branches of the gauge theory are geometrically realized only if the four-form flux in the Calabi--Yau fourfold phases are specified correctly. The conditions which the four-form flux in M-theory must satsify are rather stringent, but a large number of examples have been constructed in recent years as a by-product of the study of fluxes in F-theory \cite{Marsano:2010ix,Marsano:2011nn,Braun:2011zm,Marsano:2011hv,arXiv:1109.3454,Grimm:2011fx,arXiv:1202.3138,Cvetic:2012xn,Braun:2014pva,Cvetic:2015txa}, when those fluxes are being studied via M-theory. Note that the extremal transitions we are concerned with in this paper cannot be seen directly in F-theory because the Coulomb branch of the M-theory compactification does not lift to F-theory.  Nevertheless, they are an important feature of the corresponding M-theory compactification as was noticed in the above references.

For Calabi--Yau fourfolds with conifold singularities in codimension three --- describing three-dimensional $\mathcal{N}=2$ Abelian gauge theories at low energies --- the role of background fluxes in the corresponding Calabi--Yau fourfold phases  has been pioneered in the context of F-theory in ref.~\cite{Braun:2011zm} and studied in detail in ref.~\cite{Intriligator:2012ue}. In this work, we analyze the phase structure of $SU(N)$ gauge theories arising from $A_{N-1}$ surface singularities in a Calabi--Yau fourfold. To obtain the anticipated branches of the three-dimensional $\mathcal{N}=2$ $SU(N)$ gauge theory, we find that it is again essential to determine the correct four-form fluxes in the associated Calabi--Yau fourfold phases. In this work the focus is on $SU(N)$ gauge theories obtained from a twisted dimensional reduction on the surface $S$ with a trivial prinicpal $SU(N)$ bundle. Then the Coulomb branch of the gauge theory arises from the resolved Calabi--Yau fourfold phase in the absence of background flux, while the Higgs branch requires a specific non-trivial four-form flux. In this phase, the flux induces a superpotential which obstructs some of the Calabi--Yau fourfold moduli. These obstructions become essential because they allow us to identify the Higgs branch moduli space of the gauge theory with the unobstructed moduli space of the Calabi--Yau fourfold. Thus, while the four-form flux of the Higgs branch must fulfill the aforementioned consistency conditions, the flux is further constrained by the requirement that the Higgs branch moduli space arises from the flat directions of the flux-induced superpotential in the associated M-theory Calabi--Yau fourfold phase.

In order to explicitly check the anticipated interplay between phase transitions among gauge theory branches and their realizations as Calabi--Yau fourfolds, it is necessary to establish geometric tools to efficiently study the extremal transitions among the relevant Calabi--Yau geometries. The work of Mavlyutov~\cite{MR2092771,MR2169828} provides a mathematical framework to describe explicit examples, in which both the resolved and deformed Calabi--Yau fourfold phases are constructed as hypersurfaces and complete intersections in toric varieties, respectively. Analyzing this large class of examples, we demonstrate the anticipated agreement with the gauge theory predictions.

The organization of this work is as follows. In Section~\ref{sec:MandG} we review the role of four-form background fluxes for M-theory compactified on Calabi--Yau fourfolds. In Section~\ref{sec:FT} we perform the twisted dimensional reduction of  the $\mathcal{N}=1$ seven-dimensional $SU(N)$ gauge theory on the surface $S$ with trivial $SU(N)$-principal bundle. For the resulting three-dimensional $\mathcal{N}=2$ gauge theory, we deduce the spectrum and predict the geometry of the Coulumb and Higgs branch moduli spaces. In Section~\ref{sec:su2gen} the analysis of the gauge theory of the previous section is continued, emphasizing the M-theory compactification point of view and deducing some general geometric properties of the resolved and deformed Calabi--Yau fourfold phases. In Section~\ref{sec:wpex} we construct two explicit examples based upon Calabi--Yau fourfold hypersurfaces with $A_{N-1}$ surface singularities in weighted projective spaces. We construct both the resolved and the deformed Calabi--Yau fourfold phases in detail and verify the gauge theory predictions. In Section~\ref{sec:toric} we generalize these examples to hypersufaces with $A_{N-1}$~surface singularities embedded in toric varieties. For this large class of examples, we again find agreement with the gauge theory predictions. In Section~\ref{sec:conc} we present our conclusions.

\section{M-theory and G-flux} \label{sec:MandG}
The eleven-dimensional $\mathcal{N}=1$ gravity multiplet in the supergravity limit of M-theory consists of the graviton and the anti-symmetric three-form tensor field as its bosonic degrees of freedom.  The expectation value of the field strength of the three-form tensor field is known as the four-form flux~$G$. On a topologically non-trivial eleven-dimensional Lorentzian manifold~$M_{11}$ a consistently quantized four-form flux~$G$ fulfills the quantization condition \cite{Witten:1996md}
\begin{equation}
  \frac{G}{2\pi} + \frac{p_1(M_{11})}4 \,=\, H^4(M_{11},\mathbb{Z}) \ ,
\end{equation}
where $p_1(M_{11})$ is the first Pontryagin class of the manifold $M_{11}$.\footnote{For ease of notation we use the letter $G$ for both the four-form flux and its cohomological representative.} In this note we focus on M-theory compactifications on a compact Calabi--Yau fourfold~$X$ to three-dimensional Minkowski space $\mathbb{M}^{1,2}$. Then the quantization condition reduces to\footnote{For complex manifolds $X$ the first Pontryagin class is given by $p_1(X)=-c_2(TX \otimes \mathbb{C})$ in terms of the complexified tangent bundle $TX \otimes \mathbb{C} = T^{1,0}X \oplus \overline{T^{1,0}X}$ so that $p_1(X)=c_1(X)^2-2c_2(X)$ (with $c_1(X)=0$ for Calabi--Yau fourfolds $X$).}
\begin{equation} \label{eq:quant}
  \frac{G}{2\pi} - \frac{c_2(X)}2 \,\in\, H^4(X,\mathbb{Z}) \ .
\end{equation}
As a consequence, when the second Chern class $c_2(X)$ of the Calabi--Yau fourfold $X$ is not divisible by $2$, a consistent M-theory realization on the Calabi--Yau $X$ requires a non-zero and half-integral background flux $G$. Furthermore, due to the compactness of $X$ the Gauss law for the flux $G$ demands the tadpole cancellation condition \cite{Becker:1996gj,Witten:1996md,Sethi:1996es}
\begin{equation} \label{eq:tadpole}
  M \,=\, \frac{\chi(X)}{24} - \frac12 \int_X \frac{G}{2\pi} \wedge\frac{G}{2\pi}  \ ,
\end{equation}
in terms of the Euler characteristic $\chi(X)$ of the fourfold $X$ and an integer $M$ which enumerates the net number of space-time filling (anti-)M2-branes. Note that the quantization condition \eqref{eq:quant} ensures that the right hand side of the tadpole condition is always integral \cite{Witten:1996md}. In particular,  a Calabi--Yau fourfold with an even second Chern class $c_2(X)$ admits an M-theory background with vanishing four-form flux $G$, because the evenness of $c_2(X)$ geometrically implies that the Euler characteristic $\chi(X)$ of such a Calabi--Yau fourfold $X$ is divisible by $24$, c.f., ref.~\cite{Witten:1996md}.

In this note we analyze the phase structure of M-theory arising from extremal transitions of Calabi--Yau fourfolds along $A_{N-1}$~surface singularities. That is, we consider a singular Calabi--Yau fourfold~$X_0$ with an $A_{N-1}$~singularity along a smooth surface~$S$, and  we assume that $X_0$ admits a geometric transition to both a deformed Calabi--Yau fourfold $X^\flat$ and a resolved Calabi--Yau fourfold $X^\sharp$. 

In the context of M-theory compactifications extremal transitions among Calabi--Yau fourfolds are not automatically in accord with both the tadpole and the quantization condition \cite{Braun:2011zm,Intriligator:2012ue}. Thus for a M-theory transition between $X^\flat$ and $X^\sharp$ we must specify the number of space-time filling M2-branes $M^\sharp$ and $M^\flat$ and the background four-form fluxes $G^\sharp$ and $G^\flat$ in the respective Calabi--Yau fourfold phases. As in ref.~\cite{Intriligator:2012ue}, we consider phase transitions with a constant number of space-time filling M2-branes, i.e.,
\begin{equation} \label{eq:Mconst}
  M^\sharp \,=\, M^\flat \ ,
\end{equation}  
which we assume are located far from the transition. With this assumption the physics of the transition is governed by the degrees of freedom arising in the vicinity of the surface singularity of the Calabi--Yau fourfold~$X_0$, and the tadpole cancellation condition yields the transition condition
\begin{equation} \label{eq:tadcon}
   \frac{\chi(X^\flat)}{24}-\frac{\chi(X^\sharp)}{24}
   \,=\, \frac12 \int_{X^\flat} \frac{G^\flat}{2\pi} \wedge\frac{G^\flat}{2\pi} - \frac12 \int_{X^\sharp} \frac{G^\sharp}{2\pi} \wedge\frac{G^\sharp}{2\pi} \ ,
\end{equation}
where the left-hand side is solely determined by the topological change between between the transition fourfolds $X^\flat$ and $X^\sharp$. 

Choosing the right-hand side of \eqref{eq:tadcon} --- combined with the quantization condition~\eqref{eq:quant}  --- ensures that a M-theory phase transition is in accord with known anomaly cancellation conditions. It, however, does not guarantee that a transition can actually occurs dynamically. As the M-theory background fluxes generate a flux-induced superpotential $W$ and/or a twisted superpotentials $\widetilde W$ \cite{Gukov:1999ya}
\begin{equation} \label{eq:W}
  W \,=\, \int_X \Omega \wedge \frac{G}{2\pi} \ , \qquad \widetilde W \,=\, \int_X J\wedge J \wedge \frac{G}{2\pi} \ ,
\end{equation}  
an unobstructed extremal transition in M-theory is only realized along a flat direction of the flux-induced scalar potential $V$, which is a function of these flux-induced superpotentials. Here $\Omega$ is the holomorphic four form and $J$ is the K\"ahler form of the Calabi--Yau fourfold $X$. 

A simple solution for a dynamical M-theory transition is realized by a vanishing flux $G^\sharp$ and a non-vanishing primitive flux $G^\flat$, i.e., 
\begin{equation} \label{eq:Gsimple}
  G^\flat \,\ne\, 0 \ \text{with}\ G^\flat\wedge J=0 \ , \qquad G^\sharp \,=\,0 \ ,
\end{equation}
provided that the quantization condition~\eqref{eq:quant} for both
$X^\sharp$ and $X^\flat$ as well as the tadpole relation~\eqref{eq:tadcon} are met. This solution is of particular importance to us, as it geometrically realizes the Coulomb--Higgs gauge theory transitions that we focus on in this work. On the one hand --- due to $G^\sharp=0$ --- none of the geometric M-theory moduli are obstructed in the resolved Calabi--Yau fourfold~$X^\sharp$. On the other hand, there is a flux-induced superpotential
\begin{equation} \label{eq:Wflat}
  W^\flat \,=\, \int_{X^\flat} \Omega \wedge \frac{G^\flat}{2\pi} \ ,
\end{equation}
which generates a potential for some of the complex structure moduli fields in the deformed Calabi--Yau fourfold~$X^\flat$. At low energies the massive modes of the obstructed complex structure moduli are integrated out and a flux-restricted complex structure moduli space $\mathcal{M}^\flat_\text{cs}(G^\flat) \subset\mathcal{M}^\flat_\text{cs}$ remains. Geometrically, we can think of the unobstructed complex structure moduli as those complex structure deformation, under which the flux~$G^\flat$ remains of Hodge type $(2,2)$, whereas the Hodge structure of $G^\flat$ varies with respect to the obstructed complex structure moduli. Therefore, it is the flux-restricted moduli space $\mathcal{M}^\flat_\text{cs}(G^\flat)$ that yields the flat directions of the superpotential $W^\flat$ and should thus be compared to the moduli space in the effective gauge theory description at low energies. Note that the K\"ahler moduli of the Calabi--Yau fourfold~$X^\flat$ nevertheless remain unobstructed because by the primitivity assumption of the flux~$G^\flat$ no twisted superptoential is generated.

\section{Field theory analysis} \label{sec:FT}

In this section we discuss the predictions one obtains regarding the
relevant moduli spaces from our understanding of the low-energy
physics governing degrees of freedom localized near a (complex)
codimension two singularity in M-theory.  We start with a brief review
of the results of refs.~\cite{Katz:1996ht,Klemm:1996kv} on codimension
two singularities in type~II string theory, and proceed to contrast this with
the situation in M-theory. 

\subsection{Type~IIA string theory on a Calabi--Yau threefold}\label{subsec:fourd}

The moduli space of type~II compactifications on $X^\sharp$ is identified
with the moduli space of Calabi--Yau metrics and closed antisymmetric
tensor fields on $X^\sharp$.  Following ref.~\cite{Katz:1996ht} we discuss the IIA string
in the vicinity of a face of the K\"ahler cone at
which divisors are contracted to a smooth curve $C$ of $A_{N-1}$ singularities
of genus $g$; the superconformal field theory will be singular for
suitably tuned $B$-field.

The transverse $A_{N-1}$ singularity is resolved by blowing up along
$C$.  The vanishing cycles are described by a chain of $N-1$ two-spheres
$\Gamma_i$ in this space, with their intersection matrix corresponding
to the Dynkin diagram of $A_{N-1}$.  As we move about $C$ these
spheres sweep out $N{-}1$ divisors $E_i$ on $M$.
In homology there are then $(N{-}1)$ shrinking
two-cycles $\Gamma_i$, and $N{-}1$ shrinking four-cycles $E_i$.
The light soliton states are given by D2-branes wrapping chains of the
form $\Gamma_i\cup\Gamma_{i+1}\cup\cdots\cup\Gamma_j$ (with both
orientations) and under the Ramond--Ramond gauge symmetry associated to
$E_i$ their charges fill out the roots of $A_{N-1}$.

In the limit in which $C$ is large (deep in the
associated face) we can approximately think of the low-energy theory
as a twisted compactification on $C$ of the six-dimensional theory
obtained by including the massless solitons.

In flat space the six-dimensional theory contains a vector $V_M$, two
complex scalars $\phi$, and two fermions, all in the adjoint
representation of $SU(N)$.  The charged components are the soliton
states; the neutral components are supplied by the 
moduli of the ALE space.  The fields transform under a global
$SU(2)\times SU(2)$ $R$-symmetry.  The compactification breaks the
local Lorentz group as $SO(6)\to SO(4)\times U(1)\sim SU(2)\times
SU(2)\times U(1)$, and the requisite spin was determined in ref.~\cite{Katz:1996ht}
to be the identification of
\begin{align}\label{eq:twist}
J' =
J_L-J_3^{(1)}-J_3^{(2)}\ ,
\end{align}
as the generator of rotations in tangent space $TC$, where $J_L$ is the ``standard''
Lorentz generator, and the other two correspond to the Cartan elements
of the $SU(2)$ factors.  In four dimensions this leads to a theory
with $\mathcal{N}=2$ supersymmetry, $SU(N)$ gauge symmetry, and $g$ massless
hypermultiplets in the adjoint representation of the gauge group.
For $g>1$ the theory is IR free and a classical description is
reliable in the vicinity of the singular locus; for a discussion of
the special cases $g\le 1$, c.f., ref.~\cite{Katz:1996ht}.

The local structure of the moduli space near the singularity is
modeled --- deep in the cone --- by the structure of the space of vacua of
this gauge theory, which leads to the following predictions:\\
K\" ahler deformations away from the singular locus (and the
associated $B$-fields) parameterize the
Coulomb branch of the theory, along which the scalar $\phi$ in the
vector multiplet acquires an expectation value, constrained
by the potential to be diagonalizable by a gauge transformation.  The
eigenvalues of $\phi$, subject to the tracelessness condition, form 
coordinates on a $S_N$ cover of the Coulomb branch, on which the
Weyl group~$\mathcal{W}(SU(N))\simeq S_N$ 
acts via the $N-1$ dimensional representation.  At generic points
--- corresponding to smooth $X^\sharp$ --- the unbroken gauge symmetry is
$U(1)^{N-1}$, and the massless matter comprises $g(N-1)$ neutral
hypermultiplets. The Weyl group acts on these as well via $g$ copies
of the same representation, so locally the moduli space of $X^\sharp$
is a {\sl quotient\/} of a product of special K\"ahler manifolds.

In addition, the gauge theory has a Higgs branch in which the
hypermultiplets acquire nonzero expectation values. In terms of the
compactification on $X^\sharp$ this is the condensation of solitonic
states \cite{Greene:1995hu}.
This describes the deformations of $X^\flat$ smoothing the
singularity.  At generic points on this branch the gauge group is completely
broken and the (quaternionic) dimension of the Higgs branch is 
\begin{align}\label{eq:ehigs}
 \dim_\mathbb{H}{\mathcal H} \,=\, (g-1)(N^2-1)\ .
\end{align}
The Hodge numbers of the two spaces are thus related by 
\begin{align}
  h^{1,1}(X^\flat) = h^{1,1}(X^\sharp)-(N-1) \ , &&
  h^{2,1}(X^\flat) \,=\, h^{2,1}(X^\sharp) + (g-1)(N^2-1)-g(N-1)\ ,
\label{eq:hodgetrans}
\end{align}
where $g(N-1)$ is substracted in the last line of (\ref{eq:hodgetrans}) since these moduli already appear on the 
Coulomb branch, as we recall below.

We can somewhat refine this prediction.
There are special submanifolds on the Coulomb branch (meeting at the
origin) along which non-Abelian subgroups of $SU(N)$ are unbroken.  These
correspond exactly to fixed point sets of the $S_N$ action,
where eigenvalues of $\phi$ coincide. In general an unbroken symmetry
\begin{align}\label{eq:unbroken}
SU(k_1)\times\cdots\times SU(k_p)\times U(1)^{p-1}
\end{align}
where $\sum k_i = N,\,k_i\geq 1$ (and factors
of $SU(1)$ are simply to be ignored) will occur in codimension
$N-p$, and there will be $g$ massless hypermultiplets in the adjoint
representation of the unbroken group.  We can allow these to acquire
expectation values, breaking the non-Abelian part completely and
leading to a mixed branch $\mathcal{H}_{(k_1,\ldots,k_p)}$ with the
Higgs component having dimension
\begin{align}\label{eq:emix}
\dim_{\mathbb{H}}\mathcal{H}_{(k_1,\ldots,k_p)} \,=\, (g-1)\sum_{i=1}^p (k_i^2-1)+g(p-1)\ .
\end{align}

We can also see the transition to the Higgs branch along a different
path which will prove more transparent in the geometrical analysis.
At a generic point on the Coulomb branch, we can turn on expectation
values for the $g(N-1)$ neutral hypermultiplet scalars.\footnote{These are
the moduli that we alluded to immediately after eq.~\eqref{eq:hodgetrans}.}  Then, as we
tune $\phi$ to zero, the non-Abelian symmetry is not restored, and the
gauge symmetry remains $U(1)^{N-1}$.  The theory is still
IR free and we can use classical analysis.  
The hypermultiplet expectation values lead to masses for the
off-diagonal components of rank $N(N-1)$. 
Thus there are at the singular
point an additional $(g{-}1)N(N-1)$ charged hypermultiplets with
the $U(1)$ charges of $g-1$ adjoints.  When these acquire
generic expectation values the gauge symmetry is Higgsed leading back
to the Higgs branch dimension in eq.~\eqref{eq:ehigs}.  In other words,
we can rewrite the second equation in (\ref{eq:hodgetrans}) as
\begin{equation} \label{eq:h21}
  h^{2,1}(X^\flat) = h^{2,1}(X^\sharp) + (g-1)(N^2-N)-(N-1)\ ,
\end{equation}
understanding the additional moduli as arising from Higgsing the $U(1)^{N-1}$
under which the $(g{-}1)N(N-1)$ hypermultiplets are charged.

\subsection{M-theory on a Calabi--Yau fourfold}

The discrete choices determining a compactification of M-theory to
three dimensions include, as discussed above, a choice of a
topological type for the Calabi--Yau fourfold~$X$ as well as a choice of the four-form flux $G$
and the number $M$ of M2-branes, satisfying the conditions
\eqref{eq:tadpole} and \eqref{eq:quant}.  Given such a choice, the
moduli space is determined by the subspace of the space of Calabi--Yau metrics on
$X$ for which the chosen four-form flux~$G$ is of Hodge type $(2,2)$ as well
as primitive.  There are additional moduli associated to periods of
the three-form $A_3$ --- i.e., $h^{2,1}(X)$ of these ---
as well as to the positions of the M2-branes.  

As above we wish to consider a face of the K\"ahler cone of a Calabi--Yau fourfold~$X^\sharp$
at which a divisor contracts to a smooth surface $S$ of transverse $A_{N-1}$
singularities.  As above, this is resolved by blowing up $N-1$ times along $S$
producing $N-1$ exceptional divisors $E_i$ and the vanishing cycles
are the $E_i$ as well as $N-1$ two-cycles $\Gamma_i$.
We will assume here that we can make a choice of flux
on $X^\sharp$ such that generic points in the vicinity of this face
correspond to K\" ahler classes for which the flux is primitive.  This
means that there are M-theory vacua associated to a smooth  $X^\sharp$
in which the singularities have been resolved.  This implies that the flux, 
if nonzero, is primitive for smooth $X^\sharp$, meaning that --- if $J^i$ are the 
(1,1) cohomology classes dual to $E_i$ --- then $G^\sharp\wedge J^i=0$ for all $i$.  In 
the cases we consider here this very restrictive condition will be met by setting
$G^\sharp=0$.   Further, we assume
that the positions of the M2-branes are all far from the
contracting divisors, so that the worldvolume degrees of freedom
decouple from the low-energy theory of the modes at the singularity.

In this situation, we can perform a calculation of the low-energy
theory in the vicinity of the singular locus along similar lines to
those followed above.  In a suitable region (near a point deep in the
face of the K\" ahler cone) bulk
modes decuple from the dynamics near the singular locus
and the low-energy dynamics is given by a $\mathcal{N}=1$ seven-dimensional supersymmetric
Yang--Mills theory on $M^{1,2}\times {\mathbb R}^4$, in which the charged modes are excitations of
membranes wrapping vanishing cycles and the neutral modes describe the
moduli.  We then
want to perform a suitably twisted dimensional reduction of this seven-dimensional
theory, in which we replace ${\mathbb R}^4$ by a compact K\"{a}hler manifold~$S$, such
that we obtain an ${\mathcal N}=2$ theory in three dimensions.

The seven-dimensional supersymmetric Yang-Mills theory is obtained by dimensional
reduction from ten dimensions and is given in ref.~\cite{Ludeling:2011ip}.  
It has a global $SU(2)_R$ R-symmetry, and the fields are a gauge field $A_M$,  a triplet $S_i$ of scalars,
and a doublet  $\Psi_\alpha$ of gaugini  satisfying the symplectic Majorana condition 
\begin{equation}
\Psi_\alpha = \epsilon_{\alpha\beta} B{\Psi^\beta}{}^*\ ,
\end{equation}
where the complex conjugation matrix $B$ satisfies
\begin{equation}
B^{-1}\Gamma^M B = \Gamma^M{}^*\qquad B B^* = -{\mathrm {id}}\ .
\end{equation}
All the fields transform in the adjoint representation of the gauge
group.  

As noted above, the charged fields correspond to excitations
of M2-branes wrapping the collapsing cycles, while the Cartan elements
are associated to moduli of the compactification on $X^\sharp$.  The
gauge field $A$ is associated to periods of the three-form field $A_3$
along the vanishing cycles; the scalars $S_i$ are moduli of the metric
resolving the singularity.

The supersymmetry variations are parameterized by a symplectic Majorana spinor
doublet~$\epsilon_I$ and the relevant one for us is \cite{Ludeling:2011ip}
\begin{equation}
\delta\Psi_\alpha = -\frac14 F_{MN}\Gamma^{MN}\epsilon_\alpha  + \frac{i}2\Gamma^M D_M\left(S_i\sigma^i\right)_\alpha{}^\beta\epsilon_\beta + 
\frac14\epsilon^{ijk}[S_i,S_j](\sigma_k)_\alpha{}^\beta\epsilon_\beta\ .
\end{equation}

\subsubsection{Twisted dimensional reduction} \label{sec:Codim2Sing}

When we dimensionally reduce, the Lorentz group is reduced 
$SO(1,6)\times SU(2)_R\to SO(1,2)\times SO(4)\times SU(2)_R$ with the
$SO(4)$ eventually broken by the curvature of $S$.  
The representations in which the fields transform reduce as
\begin{align}
A:\ ({\bf 7},{\bf 1})&\to ({\bf 3},{\bf 1},{\bf 1}) \oplus ({\bf 1},{\bf 4}_v,{\bf 1})\ , \nonumber\\
S:\ ({\bf 1},{\bf 3})&\to ({\bf 1},{\bf 1},{\bf 3})\ ,\\
\Psi:\ ({\bf 8},{\bf 2})&\to ({\bf 2},{\bf 4}_s,{\bf 2})\nonumber\ .
\end{align}

Since $S$ is K\"{a}hler, the structure group is in fact $U(2)\sim
SU(2)_L\times U(1)_I$, under
which
\begin{align}
{\bf 4}_v\to {\bf 2}_1\oplus{\bf 2}_{-1} \ , \qquad
{\bf 4}_s\to {\bf 2}_0\oplus{\bf 1}_1\oplus{\bf 1}_{-1} \ .
\end{align}

\def\qf{\mathtt{q}}
In a twisted reduction, we will replace $U(1)_I$ generated by $J_I$ with $U(1)'$ generated 
by the linear combination $J' = J_I + 2 J^3_R$.  The curvature of $S$ will then couple to the
twisted $U(2)$ and the unbroken global symmetry will be $U(1)_R$ generated by $2J^3_I$.
Under $SO(1,2)\times SU(2)_L\times U(1)'\times U(1)_R$ we have the decompositions
\begin{align}
({\bf 7},{\bf 1})&\to ({\bf 3},{\bf 1})_{0,0} \oplus ({\bf 1},{\bf 2})_{1,1}\oplus({\bf 1},{\bf 2})_{-1,-1}\ ,\nonumber\\
({\bf 1},{\bf 3})&\to ({\bf 1},{\bf 1})_{2,2}\oplus ({\bf 1},{\bf 1})_{-2,-2}\oplus ({\bf 1},{\bf 1})_{0,0}\ ,\\
({\bf 8},{\bf 2})&\to ({\bf 2},{\bf 2})_{1,1}\oplus  ({\bf 2},{\bf 2})_{-1,-1}\oplus ({\bf 2},{\bf 1})_{2,1} \oplus ({\bf 2},{\bf 1})_{0,-1}\oplus ({\bf 2},{\bf 1})_{0,1}\oplus ({\bf 2},{\bf 1})_{-2,-1}\ .\nonumber
\end{align}
We identify the corresponding modes of the fields by their
transformation under $SU(2)_L\times U(1)'$ as 
\begin{align}
A_M\to A_\mu, \overline A_{\overline m},A_m\ , \qquad
S_i\to \qf,\overline \qf \ , \qquad
\Psi \to \overline\psi_{\overline m},\psi_{m},\chi,\lambda_-,\lambda_+,\overline\chi\ .
\end{align}

Our model for the local moduli space will be the space of
supersymmetric vacua of this theory.  Following ref.~\cite{Beasley:2008dc}, 
we will construct this by evaluating the supersymmetry variation of $\Psi_+$ under
the two unbroken supersymmetries. Setting this to zero 
yields a slight modification of the Vafa--Witten equations \cite{Vafa:1994tf}
\begin{equation}\label{eq:VWJ}
\begin{aligned}
&F^{(2,0)} \,=\, 0 \ , \qquad
J\wedge F^{(1,1)} + [\qf,\overline \qf] \,=\, 0 \ ,\qquad [\qf,\Phi] \,=\, 0 \ , \\
&D\Phi= \overline D\Phi \,=\,  0 \ , \qquad \overline D \qf \,=\, 0\ ;
\end{aligned}
\end{equation}
setting the variation of $\Psi_-$ to zero yields the complex conjugate
equations by the symplectic Majorana condition.

\subsubsection{Predictions for the moduli space} \label{sec:ModSpace}

Solutions to these equations provide our predictions for the local
structure of the moduli space.  Clearly the space of solutions breaks
up into disjoint components labeled by the Chern classes of the flux~$F$.
In the situation we are describing, in which the generic point in the
space of compactifications on $X^\sharp$ near the singular locus is
smooth, the bundle we obtain will be flat.  The charged components of
the curvature are certainly zero away from the singular locus since
they are carried by wrapped branes which become massive; the neutral
components can be described as the integrals 
of $G^\sharp$ over the fibers $\Gamma_i$, which we are assuming vanish.  

Three dimensional ${\mathcal N}=2$ gauge theories contain additional discrete
parameters, supersymmetric Chern--Simons couplings for the gauge fields.   
In general, the effective gauge theory describing the low-energy physics 
near a singularity will have nonzero Chern--Simons couplings and these can 
have different values between the Coulomb and Higgs vacua \cite{Intriligator:2012ue}.  
In the cases we discuss here, these couplings vanish on the Coulomb branch and 
the fact that all of the chiral multiplets are in real representations means this will
also be the case on the Higgs branch.

Linearizing about a trivial $SU(N)$ principal bundle, the modes of
$(A_\mu,\lambda_\pm,\Phi)$ form vector multiplets with masses
associated to eigenvalues of the Laplacian on $S$.  The modes of
$(A_m,\psi_m)$ form chiral multiplets with masses associated to
eigenvalues of the Laplacian on $(0,1)$-forms on $S$.  The modes of
$(\qf,\chi)$ form chiral multiplets with masses associated to
eigenvalues of the Laplacian on $(2,0)$-forms on $S$.  The massless
modes will thus be $h^0(S)=1$ vector multiplet
and $h^{0,1}(S)+h^{2,0}(S)$ chiral multiplets, all in the adjoint
representation of the gauge group.   
For the multiplicities of the chiral multiplets, we will
also use the irregularity~$q$ and the geometric genus~$p_g$,
which are respectively the conventional birational invariants for the dimensions
$h^{1,0}(S)=h^{0,1}(S)$ and $h^{2,0}(S)=h^0(K_S)$ of the algebraic surface $S$.

We can then write the low-lying excitations in terms of a basis $e_i$
for $H^{0,1}(S)$, a basis $E_A$ for $H^0(K_S)$ and the dual basis
$E^B$ for $H^0(\overline K_S)$, i.e., 
\begin{equation}
\Phi \,=\, \phi  \ , \qquad 
A_m \,=\, \sum_i a^i e_i \ , \qquad
\qf \,=\, \sum_A \qf^A E_A \ , \qquad
\overline \qf \,=\, \sum_B \overline \qf_B \overline E^B\ ,
\end{equation}
with the three-dimensional fields, $\phi$, $a^i$, $\qf_A$, $\overline \qf^B$,
taking values in the Lie algebra and $\phi$ is real. In terms of these the
conditions for unbroken supersymmetry reduce to
\begin{equation}
\begin{aligned}
&[a^i,\phi] \,=\, 0 \ , \qquad
\sum_{j,A} C_{ijA}[a^j,\qf^A] \,=\, 0 \ , \qquad
[\qf^A,\phi] \,=\, 0 \ , \\
&[\overline \qf_B,\phi] \,=\, 0\ , \qquad
\sum_A [\qf^A,\overline \qf_A] \,=\, 0\ , 
\end{aligned}
\end{equation}
where
\begin{align}
C_{ijA} = \int_S e_i\wedge e_j\wedge E_ .
\end{align}
The superpotential that leads to these equations has the form 
\begin{align}\label{eq:Yukawa}
W = \sum_{i,j,A} C_{ijA}\,{\mathrm Tr}\left([a^i,a^j]  \qf^A\right) \ .
\end{align}

In the following we restrict our analysis to surfaces $S$ with $q=0$ and
$p_g\ge 1$. As we will see in the following, this assumption ensures that there are no
non-perturbative corrections to the Coulomb branch of the gauge theory.

The model then has a Coulomb branch along which the real scalar $\phi$
acquires an expectation value. By a gauge transformation this can be
taken to lie in the Cartan algebra.  As in the previous subsection we can
use the first $N-1$ eigenvalues as coordinates, and the Weyl
group~$\mathcal{W}(SU(N)) \simeq S_N$ acts on
these.   At generic points the gauge symmetry is $U(1)^{N-1}$.
In fact \cite {Intriligator:2013lca} the Coulomb branch is complex
K\"ahler.  The periods of the dual six-form $A_6$ along $E_i$ are neutral
scalars dual to the neutral gauge bosons, and they combine with $\phi_i$ to form
holomorphic coordinates.
At a generic point the massless modes of $\qf^A$ are those
commuting with $\phi$ so we have $p_g(N-1)$ neutral massless
chiral mutiplets.  Again, the Weyl group $S_N$ acts on these and the moduli space is
a quotient.

There is another branch of the moduli space, along which $\qf^A$ acquires
a nonzero expectation value, completely breaking the gauge group and 
accordingly $\phi$ becomes massive.  The
last equation of eqs.~\eqref{eq:VWJ} is the moment map for the adjoint action of the
gauge group leading as usual to a Higgs branch of complex dimension
\begin{align}\label{eq:HBranch}
 \dim_\mathbb{C}{\mathcal H} \,=\, (p_g-1)(N^2-1)\ .
\end{align}
  This (Higgs) branch is interpreted as a local model for
the moduli space of the compactification on $X^\flat$, but we
will in general find nonzero flux. Thus (\ref{eq:HBranch})
does not lead directly to a
prediction for the deformation space of the moduli of $X^\flat$. 

As above, we have the more refined picture of the way these branches
intersect at singular loci.  At the codimension $N-p$ locus in the 
Coulomb branch along which the unbroken gauge group is given by 
eq.~\eqref{eq:unbroken} we will have $p_g$ massless chiral
multiplets in the adjoint representation.  Turning on generic expectation
values for $\qf^A$ breaks the non-Abelian part completely and
leads to a mixed branch $\mathcal{H}_{(k_1,\ldots,k_p)}$ with the Higgs
component having dimension
\begin{align}\label{eq:emixM}
  \dim_\mathbb{C}\mathcal{H}_{(k_1,\ldots,k_p)} \,=\, (p_g-1)\sum_{i=1}^p (k_i^2-1)+p_g(p-1)\ .
\end{align}

As in the previous subsection, there is another path in moduli space
implementing the transition from the Coulomb branch to the Higgs
branch.  At a generic point on the Coulomb branch we can turn on
expectation values for the $p_g(N-1)$ neutral chiral fields.  Then,
as we tune $\phi$ to zero the gauge symmetry remains $U(1)^{N-1}$.  The
D-term condition  --- implementing the third equation of eqs.~\eqref{eq:VWJ} --- then leads to
a mass term for the charged chiral fields leaving 
$(p_g-1)N(N-1)$ charged fields with the charges of $p_g-1$ adjoints, 
leading as was found in ref.~\cite{Intriligator:2012ue} to a Higgs branch 
along which the gauge symmetry is completely broken and whose 
dimension agrees with our calculation above.

\subsubsection{Quantum corrections and region of validity}

Our discussion above has been entirely classical, and we need to address
the degree to which quantum corrections might invalidate our conclusions.
We are using the effective three dimensional field theory to make predictions
about the moduli space of M-theory compactifications, and relating this to the
moduli space of K\"ahler and complex structure deformations of the Calabi--Yau
fourfolds $X^\sharp$ and 
$X^\flat$, respectively.  We thus need to check separately the degree to which 
our description of the space of vacua is subject to corrections from non-trivial
low-energy dynamics in the three-dimensional gauge theory, and the degree to which the
geometric moduli space agrees with the space of M-theory compactifications.
In the examples of Section~\ref{subsec:fourd} the four dimensional gauge 
theory was (for $g>1$) IR free and semiclassical considerations provided an 
accurate description of the moduli space near the origin.   The Coulomb
branch suffered no string corrections, and $\alpha'$ corrections to the 
metric were computable using mirror symmetry or by taking advantage of
special geometry to relate them to the holomorphic prepotential; the 
relevant parts of this, deep in the singular cone, were explicitly computed in 
ref.~\cite{Katz:1996ht}.  On the Higgs branch, $\alpha'$ corrections were absent 
and non-perturbative string effects were suppressed deep in the cone.

In the case at hand, the first new phenomenon is that the encountered three-dimensional 
gauge theories are strongly coupled at low energy.  The low-energy
dynamics of ${\mathcal N}=2$ gauge theories in three dimensions has been 
studied, for example in
refs.~\cite{Affleck:1982as,Aharony:1997bx,deBoer:1997kr,Diaconescu:1998ua,Kapustin:1999ha,
Dorey:1999rb,Tong:2000ky,Aganagic:2001uw,Intriligator:2012ue,Intriligator:2013lca}.
At the origin, we find non-trivial
interacting superconformal field theories for $p_g>1$.    For $p_g=1$ the low-energy supersymmetry
is enhanced to $\mathcal{N}=4$.  The metric on the Coulomb branch is 
subject to both perturbative and non-perturbative corrections.
For $p_g\ge 1$ the singularity at the origin of the Coulomb branch is unchanged by these, 
and the Higgs branch has the predicted singular structure \cite{deBoer:1997kr}.

The moduli space of M-theory compactifications on $X$ 
is the space of Calabi--Yau metrics and $A_3$ periods subject to the
conditions imposed by $G$-flux. The metric is subject to corrections but
the superpotential is corrected only by five-brane instantons \cite{Witten:1996bn}. 
Deep in the cone we expect the contributions of these to be suppressed, with the
exception of the contributions of five-branes wrapping the vanishing divisors $E_i$
in the case of the Calabi--Yau fourfold~$X^\sharp$. In the
absence of background flux~$G^\sharp$,
these can contribute to the superpotential only if 
$\chi(E_i,{\mathcal O}_{E_i}) = 1$.  But in the case
at hand, the divisors $E_i$ are $\mathbb{P}^1$-fibrations over the surface~$S$,
and therefore we find $\chi(E_i,{\mathcal O}_{E_i}) = p_g - q + 1$.   In the 
cases studied here, where $q=0$ and $p_g>0$, these instantons cannot contribute to the
superpotential. Therefore, the classical description of the moduli space remains
valid.  The absence of a non-perturbative superpotential can be taken as another confirmed
prediction of our identification of the gauge theory at the singularity.\footnote{See
ref.~\cite{Diaconescu:1998ua}, for a calculation of the non-perturbative superpotential in a situation where it is nonzero.}

\section{General features of $SU(N)$ models} \label{sec:su2gen}
In this section we describe some general geometric properties of the $SU(N)$ gauge theories studied in the previous section. In particular, we establish how M-theory compactified on Calabi--Yau fourfolds realizes the phase structure of the $SU(N)$ gauge theory.

\subsection{Gauge theories from surface singularities in Calabi--Yau fourfolds}
To realize geometrically the twisted dimensional reduction along the surface $S$, let us consider a Calabi--Yau fourfold~$X_0$ with a smooth surface $S$ of $A_{N-1}$ singularities. We assume for simplicity of exposition
that a tubular neighborhood of the surface $S$ in $X_0$ is given by the hypersurface equation
\begin{equation} \label{eq:singeq}
  xy = z^N  \ ,
\end{equation}
in the total space of the bundle $\mathcal{L}_1\oplus\mathcal{L}_2\oplus\mathcal{K}_S$ with $x$, $y$ and $z$ sections of the bundles $\mathcal{L}_1$, $\mathcal{L}_2$ and the canonical line bundle $K_S$, respectively. We further assume that the canonical line bundle~$K_S$ is sufficiently ample.

M-theory compactified on the singular Calabi--Yau fourfold $X_0$ yields the twisted dimensional reduction along $S$ studied in Section~\ref{sec:Codim2Sing}. That is to say that the eleven-dimensional supercharge $Q_{11}$ dimensionally reduces on the surface $S$ to the seven-dimensional supercharge $Q_7$, which --- due to the origin of $S$ as a subvariety in the ambient space $X_0$ with trivial canonical class --- becomes a section of the canonical spin$^c$ bundle~$\mathcal{S}_S^c$. This canonical spin$^c$ bundle arises from a spin$^c$~structure on $S$ associated to a spin structure on $TS\oplus K_S$. Therefore --- assuming first that the surface $S$ has a spin structure --- the twisted dimensional reduction along $S$ amounts to tensoring the spin bundle $\mathcal{S}_S$ of the surface $S$ with $K_S^{1/2}$, namely
\begin{equation}
  \mathcal{S}_S \, \xrightarrow[\text{twist}]{\otimes K_S^{1/2}} \, \mathcal{S}_S \otimes K_S^{1/2} =  \mathcal{S}_S^c \ .
\end{equation}
But even if the surface $S$ is not spin --- i.e., both the spin bundle $\mathcal{S}_S$ and the square root of $K_S^{1/2}$ are simultaneously ill-defined --- we can still formally perform the twist by tensoring with $K_S^{1/2}$, because there exists always a canonical spin$^c$~structure on $S$ such that the resulting spin$^c$ bundle $\mathcal{S}_S^c$ is well-defined.\footnote{A non-vanishing second Stiefel--Whitney class $w_2(S)\in H^2(S,\mathbb{Z}_2)$ is the obstruction to the existence of a spin structure on $S$. As $w_2(S)=w_2(K_S)$, it is also the obstruction to the existence of a square root of the canonical line bundle. Therefore, we have $w_2(TS\oplus K_S) = w_2(S) + w_2(K_S) = 0$, which implies the existence of a canonical spin$^c$~structure on $S$.}

Since the twist acts on the supercharges, which generate the resulting spectrum of three-dimensional $\mathcal{N}=2$ supermultiplets, tensoring with $K_S^{1/2}$ realizes the twist of $J_I$ as in Section~\ref{sec:Codim2Sing}. This yields geometrically the previously determined three-dimensional $\mathcal{N}=2$ supersymmetric Yang--Mills spectrum of a single vector multiplet and $p_g+q$ chiral matter multiplets in the adjoint representation of $SU(N)$.

\subsection{The change in topology for the Coulomb--Higgs phase transition}
In the M-theory compactification the Coulomb branch of the gauge theory realizes a crepant resolution of the singular Calabi--Yau fourfold $X_0$ to the resolved Calabi--Yau fourfold $X^\sharp$. This amounts to replacing the $A_{N-1}$~surface singularity along $S$ in $X_0$ by a chain of $N$ $\mathbb{P}^1$-bundles over $S$.

The Higgs branch of the gauge theory describes deformations of the singular Calabi--Yau fourfold $X_0$ to the Calabi--Yau fourfold $X^\flat$, locally given by the deformed hypersurface equation
\begin{equation} \label{eq:ANDef}
  xy = z^N + \sum_{j=0}^N \omega_{N-j} z^j \ .
\end{equation}
Here $\omega_j$ are sections of the pluri-canonical line bundles $j K_S$. 

To determine the change in Euler characteristic for the transition, we compare the smooth Calabi--Yau fourfold $X^\sharp$ to the smooth Calabi--Yau fourfold $X^\flat$. Let us specialize to the fourfold~$X^\flat$ arising from $\omega_j=0$ for $j<N$ but with a generic section $\omega_N$ of $N\,K_S$, such that the equation~\eqref{eq:ANDef} becomes $xy = z^N + \omega_N$. Then the $A_{N-1}$ surface singularity in $X^\flat$ is replaced by a bundle of a bouquet $S^2 \vee \ldots \vee S^2$ of $N-1$ two-spheres collapsing over the curve $\mathcal{C} \subset S$, where $\mathcal{C}$ is the vanishing locus of $\omega_{N}$. We assume that for the generic choice of $\omega_{N}$ the curve $\mathcal{C}$ is smooth.\footnote{If the generic curve $\mathcal{C}$ is not smooth, additional massless matter fields are present in the gauge theory spectrum.} Thus, as both smooth Calabi--Yau fourfold phases $X^\sharp$ and $X^\flat$ arise from topological fibrations of a bouquet of $N-1$ two-spheres over $S\setminus \mathcal{C}$, the change in Euler characteristic between $X^\sharp$ and $X^\flat$ is determined by the difference in Euler characteristic of the fibrations along the curve $\mathcal{C}$. The Euler characteristic of the bouquet of $N$ two-spheres fibered over $\mathcal{C}$ in $X^\sharp$ becomes $N\cdot\chi(\mathcal{C})$ and compares to the Euler characteristic $1\cdot\chi(\mathcal{C})$ of the collapsed fibers over $\mathcal{C}$ in $X^\flat$, so that
\begin{equation} \label{eq:Euler1}
   \chi(X^\sharp) - \chi(X^\flat) \,=\, (N-1)\chi(\mathcal{C}) \ .
\end{equation}   
The Euler characteristic $\chi(\mathcal{C})$ in turn is minus the degree of the canonical bundle $K_{\mathcal{C}}$ of $\mathcal{C}$, which by the adjunction formula is computed to be $K_\mathcal{C} = (N+1) K_S |_{\mathcal{C}}$. Therefore, the canonical bundle $K_\mathcal{C}$  has degree $(N+1)N\,K_S^2$, and we arrive with eq.~\eqref{eq:Euler1} at
\begin{equation} \label{eq:Euler2}
\chi(X^\flat) -  \chi(X^\sharp)  \,=\, N (N-1)(N+1) K_S^2 \ .
\end{equation}  
This argument generalizes from the $A_{N-1}$ case to Calabi--Yau fourfolds $X_0$ with a smooth surface $S$ of ADE~singularities. Then the curve $\mathcal{C}\subset S$ becomes the vanishing locus of a section of the line bundle $h_G K_S$ in terms of the dual Coxeter number $h_G$ of the ADE group $G$, and we obtain
\begin{equation}
  \chi(X^\flat) -  \chi(X^\sharp)   \,=\, r_G h_G(h_G + 1) K_S^2 \ ,
\end{equation}
where $r_G$ is the rank of the group $G$.

\subsection{The gauge theory and geometric moduli space}
Before we study explicit examples in the next section, we make some further general remarks about the relationship between the gauge theory and the geometric moduli spaces.

In the Coulomb branch of the gauge theory the adjoint-valued scalar field $\phi$ of the vector multiplet acquires an expectation value in the Cartan subalgebra
\begin{equation} \label{eq:DMat}
   \langle \phi \rangle \,=\, \operatorname{Diag}\left( \phi_1 , \ldots , \phi_{N} \right) \ , \qquad \phi_1 + \ldots + \phi_{N} = 0 \ ,
\end{equation}
which generically breaks the gauge group $SU(N)$ to its maximal Abelian subgroup $U(1)^{N-1}$. The Weyl group $\mathcal{W}(SU(N))\simeq S_{N}$ of $SU(N)$ permutes the expectation values $\phi_j$, $j=1,\ldots, N$. Hence, we can view the expectation values $\phi_j$ as coordinates on the $S_{N}$-covering space of the $N-1$-dimensional moduli space of the Coulomb branch. To describe the Coulomb moduli space itself --- and not its $N!$-fold cover --- we pick in the Weyl orbit of diagonal expectation values~\eqref{eq:DMat} a representative obeying
\begin{equation} \label{eq:PhiOrdered}
  \phi_1 \ge \phi_2 \ge \ldots \ge \phi_{N} \ .
\end{equation}
In the Calabi--Yau fourfold $X^\sharp$ the non-negative differences $J^j=\phi_j - \phi_{j+1}$, $j=1,\ldots,N-1$, become K\"ahler coordinates for the $N-1$ exceptional divisors in the chain of $N-1$ $\mathbb{P}^1$ fibrations over the surface $S$. If any two expectation values in $\langle \phi \rangle$ coincide, the representative~\eqref{eq:PhiOrdered} of the Weyl orbit ceases to be unique and the gauge group is not entirely broken to the maximal Abelian subgroup. Geometrically, some of the K\"ahler moduli $J^j$ vanish, and hence we are on the boundary of the K\"ahler cone. This means that the $A_{N-1}$ surface singularity is not entirely resolved in the Calabi--Yau fourfold $X_0$, which geometrically reflects that the gauge group is only partially broken to a group properly containing its maximal Abelian subgroup.

In the Higgs branch of the gauge theory it is the adjoint-valued matter fields $\qf$ that acquire an expectation value. The expectation values of the matter fields $\qf$ are adjoint-valued sections of $H^0(S,K_S)$ and deform the ADE~surface singularity~\eqref{eq:singeq}. The deformations are governed by invariant theory of the $SU(N)$ gauge group and take the form 
\begin{equation} \label{eq:defHiggs}
   xy \,=\, \det \left( z \cdot I_N + M \right) \ .
\end{equation}
Here $M$ is a traceless $N\times N$ matrix whose entries are sections of the canonical line bundle~$K_S$. The deformed hypersurface equation~\eqref{eq:defHiggs} is manifestly Weyl invariant, as the Weyl group $S_{N}$ acts on $\qf$ by conjugation with permutation matrices
\begin{equation} \label{eq:WeylHiggs}
  \sigma \in S_{N}: \ q \mapsto P_\sigma^\dagger q P_\sigma \ .
\end{equation}

We observe that --- in agreement with the gauge theory prediction for the Higgs branch in Section~\ref{sec:ModSpace} --- the invariant deformations~\eqref{eq:defHiggs} parametrize a $(N^2-1)\cdot (p_g-1)$-dimensional subspace in the space of all hypersurface deformation~\eqref{eq:ANDef}. In the M-theory compactification on the Calabi--Yau fourfold~$X^\flat$, the gauge invariant deformations~\eqref{eq:defHiggs} become the flat directions of the superpotential~\eqref{eq:Wflat} arising from a suitable four-form flux $G^\flat$. The flux $G^\flat$ is required to be primitive, to fulfill the quantization condition~\eqref{eq:quant}, and to accommodate for the tadpole cancellation condition~\eqref{eq:tadcon}, which implies together with eq.~\eqref{eq:Euler2} that
\begin{equation} \label{eq:fluxcont}
  \frac12 \int_{X^\flat} \frac{G^\flat}{2\pi} \wedge\frac{G^\flat}{2\pi} \,=\, \frac{1}{24} N(N-1)(N+1) K_S^2 \ .
\end{equation}
We now claim that in the neighborhood where the Calabi--Yau fourfold $X^\flat$ is described in terms of the hypersurface equation~\eqref{eq:defHiggs} the flux $G^\flat$ is locally given by
\begin{equation} \label{eq:Gflat}
   \frac{G^\flat}{2\pi} \,=\, \frac{N-1}{2} R - T \ .
\end{equation}
Here $R$ is the two-dimensional algebraic cycle arising from the intersection
\begin{equation}
    R: \quad x\,=\,z\,=\,0  \ ,
\end{equation}
while $T$ is the two-dimensional algebraic cycle given by
\begin{equation}
   T: \quad x\,=\,0 \ , \ \operatorname{rank} S \,\le\, N-2 \ ,
\end{equation}
in terms of the $N\times (N-1)$ submatrix $S$ of the $N\times N$ matrix $z \cdot I_N + M$ obtained by deleting the last column. Note that the construction of the submatrix $S$ is not gauge invariant, as gauge transformations act upon the matrix $z\cdot I_n+M$ by conjugation. As a matter of fact there is a whole Weyl orbit of algebraic cycles $R$ obtained form conjugation by permutation matrices according to eq.~\eqref{eq:WeylHiggs}, which give rise to equivalent flux-restricted moduli space $\mathcal{M}^\flat_\text{cs}(G^\flat)$ in agreement with the Higgs branch moduli space~\eqref{eq:HBranch}.

The detailed local analysis of Calabi--Yau fourfolds with $A_{N-1}$ singularities in codimension two together with the structure of local background fluxes is presented elsewhere~\cite{WProg}. Here we justify our proposal in the context of extremal transitions in global Calabi--Yau fourfolds. Namely, for a rather large class of toric example to be studied in the next two sections, we explicitly spell out a consistent background flux $G^\flat$, which in the vicinity of the deformed $A_{N-1}$ surface singularity agrees with our local proposal~\eqref{eq:Gflat} for the four-form flux~$G^\flat$. 

We observe that the Weyl group $\mathcal{W}(SU(N))\simeq S_N$ acts on the matrix $M$ according to eq.~\eqref{eq:WeylHiggs}, and hence induces a non-trivial action on the submatrix $S$ of the flux component $T$, generating the Weyl orbit of fluxes~$G^\flat$. While non-trivially acting on the flux~$G^\flat$, the Weyl group $S_N$ does not change any complex structure moduli because $\det \left( z \cdot I_N + M \right)$ remains invariant with respect to conjugation by permutation matrices. Therefore, the Weyl group action realizes a monodromy in the M-theory moduli space, which is fibered over the complex structure moduli space of $X^\flat$. Nevertheless, the restricted complex structure moduli space~$\mathcal{M}^\flat_\text{cs}(G^\flat)$ --- identified with the Higgs branch of the gauge theory --- remains invariant under the monodromy action upon the flux~$G^\flat$. It is rather surprising that we find a remnant of the Weyl group in the M-theory moduli space of the Calabi--Yau phase associated to the gauge theory Higgs branch. It would interesting to further study the implications of this observation. We will give an explicit example of this phenomenon in Section~\ref{sec:example}.

\section{Examples in weighted projective spaces}
\label{sec:wpex}
In this section we provide two examples of hypersurfaces in weighted projective space
to help fix ideas in a global setting:
an $SU(2)$ example (which is similar to an example previously presented in ref.~\cite{Braun:2011zm})
and an $SU(6)$ example.

\subsection{Calabi--Yau fourfolds from $\mathbb{P}^{(1,1,2,2,2,2)}$}
\label{sec:example}

\medskip
We consider a generic weighted hypersurface $X_0$ of weight 10 in
$\mathbb{P}^{(1,1,2,2,2,2)}$ defined by a weight~10 polynomial $f_{10}(x_1,\ldots,x_6)$,
its desingularization $X^\sharp$, 
and its smoothing $X^\flat$.

\subsubsection{Geometric data}
\label{subsubsec:su2geom}
Since $\mathbb{P}^{(1,1,2,2,2,2)}$ is the quotient of $\mathbb{P}^5$ by 
the $\mathbb{Z}_2$-action  
given by multiplication of the coordinates $(x_1,\ldots,x_6)$ by $(-1,-1,1,1,1,1)$, the
codimension two locus $Y$ defined by $x_1=x_2=0$ has a transverse $A_1$ singularity.  Then
$Y\simeq\mathbb{P}^3$, with $(x_3,x_4,x_5,x_6)$ serving as homogeneous coordinates.  We can 
resolve the singularity of $\mathbb{P}^{(1,1,2,2,2,2)}$ by blowing up $Y$ to get a smooth variety
$\tilde{\mathbb{P}}$.  Then the proper transform $X^\sharp$ of $X_0$ in $\tilde{\mathbb{P}}$ is smooth.

Furthermore,
$X_0$ is singular along $S=X_0\cap Y$, which has equation $f_{10}(0,0,x_3,x_4,x_5,x_6)$ in the
homogeneous coordinates of $Y\simeq\mathbb{P}^3$.  Since $x_3,\ldots,x_6$ each have weight 2,
then the weighted polynomial  $f=f_{10}(0,0,x_3,x_4,x_5,x_6)$ has degree~5 as an ordinary 
polynomial, and we have identified $S$ as a quintic hypersurface in $\mathbb{P}^3$, i.e.\ a
quintic surface.  By the Lefschetz hyperplane theorem, we have $q=h^{1,0}(S)=0$.
 
By the adjunction formula, we have  
\begin{equation}
  \label{eq:ks}
  K_S=\mathcal{O}_S(-4+5)=\mathcal{O}_S(1) \ .
\end{equation}
The number of adjoint chiral multiplets is then $p_g=h^0(K_S)=h^0(\mathcal{O}_S(1))$.  We compute this
space of sections using the exact sequence
\begin{equation}
  \label{eq:restes}
  0\to \mathcal{O}_{\mathbb{P}^3}(-4) \xrightarrow{\ \alpha\ } \mathcal{O}_{\mathbb{P}^3}(1)
\xrightarrow{\ r\ }\mathcal{O}_S(1)\to 0 \ ,
\end{equation}
where $\alpha$ is multiplication by $f$ and $r$ is restriction to $S$.  Taking cohomology
of \eqref{eq:restes} and using $H^0(\mathcal{O}_{\mathbb{P}^3}(-4))=
H^1(\mathcal{O}_{\mathbb{P}^3}(-4))=0$, we get
\begin{equation} \label{eq:pgs}
  p_g=h^0(\mathcal{O}_S(1))=h^0(\mathcal{O}_{\mathbb{P}^3}(1))=4 \ .
\end{equation}
Thus we have 4 adjoint chiral multiplets in our $SU(2)$ gauge theory.

\smallskip
To describe $X^\flat$, we first embed $\mathbb{P}^{(1,1,2,2,2,2)}$ as a singular
quadric hypersurface in $\mathbb{P}^6$ by
\begin{equation}
  \label{eq:wpembed}
\mathbb{P}^{(1,1,2,2,2,2)}\to\mathbb{P}^6 \ , \qquad (x_1,\ldots,x_6)\mapsto
(x_1^2,x_2^2,x_1x_2,x_3,x_4,x_5,x_6) \ .
\end{equation}
Letting $(y_0,\ldots, y_6)$ be homogeneous coordinates on $\mathbb{P}^6$, we
see that (\ref{eq:wpembed}) embeds $\mathbb{P}^{(1,1,2,2,2,2)}$ isomorphically onto
the singular quadric hypersurface with equation $q_0(y)=y_0y_1-y_2^2=0$.  Furthermore, after the
substitution $y_0=x_1^2$, $y_1=x_2^2$, $y_2=x_1x_2$, and $y_i=x_i$ for  $3\le i\le 6$ described
by (\ref{eq:wpembed}), we can find
a homogenous degree five polynomial $g(y)$ with $g(y)=f_{10}(x)$,  and $g(y)$ is unique up to  multiples of $q_0(y)$.  We conclude  that
$X_0$ is isomorphic to the complete intersection of $q_0(y)$ and $g(y)$,  a (singular) complete intersection Calabi--Yau  
fourfold
$\mathbb{P}^6[2,5]$.

This description makes it clear how to smooth $X_0$ to obtain $X^\flat$: simply smooth
the singular quadric $q_0(y)$ to a quadric $q^\flat(y)$ to obtain a more general complete intersection Calabi--Yau 
$\mathbb{P}^6[2,5]$.  The generic $q^\flat=0$ will intersect $g=0$ transversely,  so the
resulting Calabi--Yau fourfold will be smooth.  In fact, we can still get a smooth complete intersection Calabi--Yau if $q^\flat=0$
has an isolated singularity at which $g$ does not vanish.  
Such a $q^\flat$ is a rank~6 quadric.

To count moduli for these deformations of $q_0(y)$, the space
of first order deformations of $q_0$ is given by  the degree 2 part of
$\mathbb{C}[y_0,\ldots,y_6]/J(q_0)$.  Since the partial derivatives of $q_0$ are just $y_0,y_1,y_2$ up
to multiple, the space of first order deformations is identified with homogeneous degree
2 polynomials in $y_3,y_4,y_5,y_6$, a ten-dimensional space.

We make contact  with the discussion in Section~\ref{sec:su2gen}, where the deformation
was described by (\ref{eq:ANDef}), with $\omega_2\in H^0(S,2K_S)$.  For the quintic surface,
we have that
$H^0(S,2K_S)=H^0(S,\mathcal{O}_S(2))$.  Tensoring (\ref{eq:restes}) with $\mathcal{O}(1)$ and
using the vanishing of the cohomologies of $\mathcal{O}_{\mathbb{P}^3}(-3)$, we conclude
that $H^0(S,2K_S)$ is identified with the space of degree 2 homogeneous polynomials in 
$\mathbb{P}^3$. So the space of smoothings which we described explicitly above is canonically
identified with
$H^0(S,2K_S)$.

\subsubsection{Adding G-flux}
\label{sec:gflux}

\medskip
Letting $L\in H^2(X^\sharp)$ be the proper transform of the divisor 
$(x_1=0)\subset X_0$ and $M\in H^2(X^\sharp)$ be the proper transform of the 
divisor 
$(x_3=0)\subset X_0$, explicit computation  gives\footnote{Since this is a standard
computation, we will content ourselves with explaining how to perform an
equivalent computation in a more general context in Section~\ref{sec:toric}.}
\begin{equation}
  \label{eq:c2xs2}
  c_2(X^\sharp)=2LM+10M^2,
\end{equation}
which is visibly an even class.  Therefore $G^\sharp=0$ satisfies the
quantization condition.

We now exhibit explicit smoothings $X^\flat$ which satisfy a G-flux constraint,
parallel to a construction previously presented in ref.~\cite{Braun:2011zm}.
In the
example under investigation, $S$ is a quintic surface and
$K_S=\mathcal{O}_S(1)$ so  we have that $K_S^2=5$.
It follows according to eq.~\eqref{eq:Euler1} that the Euler characteristic changes
by $30=(N+1)N(N-1)K_S^2$ with $N=2$.  

Starting the
transition with $G^\sharp=0$ as above, 
then for $G^\flat$ we require $\frac12 \left(\frac{G^\flat}{2\pi}\right)^2=\frac{30}{24}=\frac52$.
We also require $G^\flat$ to satisfy 
the quantization condition \eqref{eq:quant} that $c_2(X^\flat)-2\cdot\frac{G^\flat}{2\pi}$ is even.

We can find a suitable $G^\flat$ after
constraining $q^\flat$ to be a rank~6 quadric (whose singular point 
$p$ is not contained in the quintic hypersurface $g=0$).  We have seen that 
we can parametrize the moduli of $q^\flat$ as
\begin{equation}
  \label{eq:tq}
  q^\flat=y_0y_1-y_2^2+\hat{q}(y_3,y_4,y_5,y_6).
\end{equation}
In this parametrization, $q^\flat$ has rank~6 if and only if $\hat{q}$ has
rank~3.  Writing 
\begin{equation}
\hat{q}=^t\!\!(y_3,\ldots,y_6)Q(y_3,\ldots,y_6)
 \end{equation}
in terms of a $4\times4$ symmetric matrix $Q$,
the condition for $\hat{q}$ to have rank~3 is that $\det Q=0$, a codimension
one condition.  This gives a $10-1=9$ dimensional moduli space, which we will
identify with the Higgs branch of the gauge 
theory after exhibiting $G^\flat$.  As a check, the dimension of the Higgs branch
of an $SU(2)$ gauge theory with 4 adjoints is $3\cdot 4-3
=9$.  

If $\hat{q}$ has rank~3, then it can be put in the form $\hat{q}=y_3^2+y_4y_5$
after a change of coordinates, leading to
\begin{equation}
  \label{eq:rank6}
  q^\flat=y_0y_1-y_2^2+y_3^2+y_4y_5.
\end{equation}
This equation can be compared with (\ref{eq:defHiggs}) by rewriting it as
\begin{equation}
  \label{eq:rewrank6}
  y_0y_1=\det\left(
y_2I_2+\left(
  \begin{array}{cc}
    y_3&y_4\\
y_5&-y_3
  \end{array}
\right)\right)
\end{equation}

Then the quadric $q^\flat=0$ contains the 3-planes $P_1$ and $P_2$
defined by
$y_0=y_3-y_2=y_4=0$ and $y_0=y_3-y_2=y_5=0$ as codimension two subvarieties.  
The 3-planes described explicitly above are in different rulings.  
However, nonsingular (rank~7) quadrics have only one ruling.

We wish to emphasize this point, which encodes the key geometric
property of our choice of flux.  A homogeneous quadric of rank $r$
in $\mathbb{P}^{r-1}$ contains an irreducible family of linear subspaces
when $r$ is odd, but contains two distinct families of linear subspaces
when $r$ is even.\footnote{This is closely related to the familiar fact that an
orthogonal group in a space of odd dimension has a single irreducible spinor 
representation, but an orthogonal group in a space of even dimension has
two distinct spinor representation.}  This statement about linear
subspaces depends only on the rank, not on the dimension in which the
quadric has been embedded.  Thus, a quadric in $\mathbb{P}^{2k}$ of
rank $2k$ has a unique singular point and two families of linear
subspaces, but when we smooth this quadric to one of rank $2k+1$ (the
generic case) there is only one family of linear subspaces.  In particular,
the difference $P_1-P_2$ of spaces from the two families exists as
a cycle on the rank $2k$ quadric which cannot be extended to a
cycle on the nonsingular quadric of rank $2k+1$.

Restricting to $X^\flat$ by intersection with $g=0$, these $P_i$ yield
codimension two subvarieties of $X^\flat$ with cohomology classes 
$T_1,T_2\in H^4(X^\flat,\mathbb{Z})$.  Since on $q^\flat$ we have
that $y_0=y_3-y_2=0$ is $P_1\cup P_2$, we have in cohomology that 
$H^2=T_1+T_2$, where $H$ is the hyperplane class of $\mathbb{P}^6$ 
restricted to $X^\flat$.
As in refs.~\cite{Braun:2011zm,Intriligator:2012ue}, we then take
\begin{equation}
  \label{eq:gflux}
  \frac{{G}^\flat}{2\pi}\,=\,\frac12\left(T_1-T_2\right) \ .
\end{equation}
Since the cycles $T_1$ and $T_2$ have the same degrees, and $H^6(X^\flat)=0$ by Lefschetz,
we see that $H\cdot G^\flat=0$ and $G^\flat$ is primitive.  Furthermore, $G^\flat$ is of 
Hodge type~$(2,2)$, as it
is an  algebraic cohomology class.   Thus all of the directions in moduli 
corresponding to rank 6 $q^\flat$ are flat directions relative to the superpotential
generated by $G^\flat$.

However, if we try to deform further to a rank~7 (nonsingular) quadric $q^\flat$, then
there is only one ruling on $q^\flat$ so we do not have cycles $T_1$ and $T_2$ in that
case.  In the absence of cycles $T_1$ and $T_2$, there is no reason for $G^\flat$ to 
remain of 
type $(2,2)$ and we expect that it is not of type $(2,2)$.  It would be interesting to verify
this expectation.
   
Returning to the situation where $q^\flat$ has rank~6, since 
each $T_i\subset\mathbb{P}^6$ is a complete intersection  of three linear forms and
a quintic,
we compute the Chern class of its normal bundle in $X^\flat$ as
\begin{equation}
  \label{eq:cn2}
  c\left(N_{T_i,X^\flat}\right)=\frac{\left(1+H\right)^3\left(1+5H\right)}{\left(1+2H\right)\left(1+5H\right)}.
\end{equation}
Expanding (\ref{eq:cn2}), we get $c_2(N_{T_i,X^\flat})=H^2$, which is numerically 5,
as $T_i$ has degree 5, owing to the intersection with $g=0$.  Thus
\begin{equation}
  T_1^2=T_2^2=5.
\end{equation}
We then compute $T_1\cdot T_2=T_1\cdot(H^2-T_1)=5-5=0$.
Thus $\left(\frac{G^\flat}{2\pi}\right)^2=\frac52$ as required.

\smallskip
For the quantization condition, explicit computation gives $c_2(X^\flat)=11H^2$.
Since we only need to compute mod 2, we can replace $2\cdot\frac{G^\flat}{2\pi}=T_1-T_2$
by $T_1+T_2=H^2$.  We learn that 
$c_2(X^\flat)-2\cdot \frac{G^\flat}{2\pi}$ is congruent mod 2 to 
$10H^2$, which is even. So the quantization condition is satisfied.

\medskip
Looking at eq.~\eqref{eq:rewrank6}, the action of the Weyl group
$\mathcal{W}(SU(2))\simeq\mathbb{Z}_2$ is realized by interchanging
rows and columns, i.e.\ 
\begin{equation}
  \label{eq:weyl2}
  \left(
  \begin{array}{cc}
    y_3&y_4\\
y_5&-y_3
  \end{array}
\right)\mapsto\left(
  \begin{array}{cc}
    -y_3&y_5\\
y_4&y_3
  \end{array}
\right),
\end{equation}
which has the effect of switching $T_1$ and $T_2$.  So the Weyl group
sends $G^\flat$ to $-G^\flat$.

Recall that $G^\flat$ is determined by a choice of ruling of a quadric, or 
equivalently, by a choice of a matrix in the representation (\ref{eq:rewrank6}) of one of the equations for $X^\flat$. Then the Weyl group action can be explicitly realized as a monodromy in
the M-theory moduli space over the complex structure moduli space.  We 
realize this monodromy as follows. Consider the space $\mathcal{M}$ of
$2\times2$ matrices of linear forms in $y_3,y_4,y_5$ whose determinant is
a rank~3 quadric.  We
choose a path in $\mathcal{M}$ which starts at the
matrix on the left of (\ref{eq:weyl2}) and ends at the matrix on the right
of (\ref{eq:weyl2}).  Explicitly, we can take
\begin{equation}  \label{eq:path}
M(\theta) \,=\, 
\left(  \begin{array}{cc}
    e^{i\theta}y_3&\frac{y_4\left(1+e^{i\theta}\right)+y_5\left(1-e^{i\theta}\right)}2 \\
\frac{y_5\left(1+e^{i\theta}\right)+y_4\left(1-e^{i\theta}\right)}2& -e^{i\theta}y_3
  \end{array}\right),\qquad 0\le\theta\le \pi \ ,
\end{equation}
so that $\theta$ parametrizes a path in the unobstructed complex structure moduli
space $\mathcal{M}^\flat_\text{cs}(G^\flat)$.  For $\theta=0$ we get $G^\flat$
as in eq.~\eqref{eq:gflux}, while for $\theta=\pi$ we get $-G^\flat$.

\subsection{Calabi--Yau fourfolds from $\mathbb{P}^{(1,5,6,6,6,6)}$}
\label{sec:su3example}

\medskip
We consider a generic weighted hypersurface $X_0$ of weight 30 in
$\mathbb{P}^{(1,5,6,6,6,6)}$ defined by a weight~30 polynomial $f_{30}(x_1,\ldots,x_6)$,
its desingularization $X^\sharp$,
and its smoothing $X^\flat$. 

\subsubsection{Geometric data}
\label{subsubsec:su6geom}

Since $\mathbb{P}^{(1,5,6,6,6,6)}$ is the quotient of $\mathbb{P}^5$ by the 
$\mathbb{Z}_6$-action  
given by multiplication of the coordinates $(x_1,\ldots,x_6)$ by $(\omega,\omega^5,1,1,1,1)$ with $\omega^6=1$ and an additional $\mathbb{Z}_5$-action, the
codimension two locus $Y$ defined by $x_1=x_2=0$ has a transverse $A_5$ singularity,
at least away from the point $p=(0,1,0,0,0,0)$ which is the isolated fixed point
of the additional $\mathbb{Z}_5$.  
Then
$Y\simeq\mathbb{P}^3$, with $(x_3,x_4,x_5,x_6)$ serving as homogeneous coordinates. 
We can 
resolve the singularity of $\mathbb{P}^{(1,5,6,6,6,6)}$ (away from $p$)
by blowing up $Y$ to get a smooth variety
$\tilde{\mathbb{P}}$.  Since a generic $f_{30}$ does not vanish at $p$, it follows that
the proper transform $X^\sharp$ of $X_0$ in $\tilde{\mathbb{P}}$ is smooth.

Furthermore,
$X_0$ is singular along $S=X_0\cap Y$, which has equation $f_{30}(0,0,x_3,x_4,x_5,x_6)$ in the
homogeneous coordinates of $Y\simeq\mathbb{P}^3$.  Since $x_3,\ldots,x_6$ each have weight 6,
then the weighted polynomial  $f=f_{10}(0,0,x_3,x_4,x_5,x_6)$ has degree~5 as an ordinary 
polynomial, and we have identified $S$ as a quintic hypersurface in $\mathbb{P}^3$, i.e., a
quintic surface.  So again we have $q=h^{1,0}(S)=0$, $K_S^2=5$ and there are 4 adjoints in our $SU(6)$ gauge theory.

\smallskip
To describe $X^\flat$, we first embed $\mathbb{P}^{(1,5,6,6,6,6)}$ as a singular
weighted hypersurface in $\mathbb{P}^{(1,5,1,1,1,1,1)}$ by
\begin{equation}
  \label{eq:wpembed6}
\mathbb{P}^{(1,5,6,6,6,6)}\to\mathbb{P}^{(1,5,1,1,1,1,1)}, \qquad 
(x_1,\ldots,x_6)\mapsto
(x_1^6,x_2^6,x_1x_2,x_3,x_4,x_5,x_6).
\end{equation}
Letting $(y_0,\ldots, y_6)$ be homogeneous coordinates on 
$\mathbb{P}^{(1,5,1,1,1,1,1)}$, we
see that (\ref{eq:wpembed6}) embeds $\mathbb{P}^{(1,5,6,6,6,6)}$ isomorphically onto
the weight~6 hypersurface with equation $q_0(y)=y_0y_1-y_2^6=0$.  
Furthermore, after the
substitution $y_0=x_1^6$, $y_1=x_2^6$, $y_2=x_1x_2$, and $y_i=x_i$ for $3\le i\le 6$
described by \eqref{eq:wpembed6}, we can find
a homogenous degree five polynomial $g(y)$ with $g(y)=f_{30}(x)$.
Note  that $g$ does not vanish at the unique singular point
$p=(0,1,0,0,0,0,0)$ of  $\mathbb{P}^{(1,5,1,1,1,1,1)}$.
We conclude  that
$X_0$ is isomorphic to the complete intersection of $q_0(y)$ and $g(y)$,  a (singular)
complete intersection Calabi--Yau  fourfold
$\mathbb{P}^{(1,5,1,1,1,1,1)}[6,5]$.

This description makes it clear how to smooth $X_0$ to obtain $X^\flat$: simply smooth
the $q_0(y)$ to a general weight~6 hypersurface $q^\flat(y)$ to obtain a more general complete intersection Calabi--Yau 
$\mathbb{P}^{(1,5,1,1,1,1,1)}[6,5]$.  The generic $q^\flat=0$ will intersect $g=0$ transversely,  so the
resulting complete intersection Calabi--Yau will be smooth.

To count moduli for these deformations of $q_0(y)$, the space
of first order deformations of $q_0$ modulo $g$ is given by  the degree 6 part of
$\mathbb{C}[y_0,\ldots,y_6]/(J(q_0),g)$.\footnote{In the case of $\mathbb{P}^{(1,1,2,2,2,2)}[10]$,
we did not have to consider the deforming polynomials modulo $g$ since the
degree of $q_0$ was less than the degree of $g$.}  Since the partial derivatives of $q_0$ are just $y_0$, $y_1$, $y_2^5$ up
to multiple, the space of first order deformations is identified with homogeneous degree
6 polynomials in $y_2,y_3,y_4,y_5,y_6$ modulo $g$, where $y_2$ occurs with degree
at most 4.  

Write these deforming polynomials as
\begin{equation}
  \sum_{j=0}^4h_{6-j}(y_3,y_4,y_5,y_6)y_2^j.
\end{equation}
Since $h_{6-j}$ is to be taken modulo $g$, we view the coefficients of $y_2^d$ as
$h_{6-d}\in H^0(S,\mathcal{O}_S(6-j))$.  So  we see that the space of smoothings
is identified with
$\oplus_{j=2}^6H^0(S,jK_S)$, in complete agreement with (\ref{eq:ANDef}).

\subsubsection{Adding G-flux}
\label{sec:gflux6}

\medskip
Resolving the $A_5$ singularity introduces five exceptional divisors, which we
denote by $E_1,\ldots,E_5$.  
Letting $L\in H^2(X^\sharp)$ be the proper transform of the divisor 
$(x_1=0)\subset X_0$, $M\in H^2(X^\sharp)$ be the proper transform of the divisor 
$(x_2=0)\subset X_0$, and $N\in H^2(X^\sharp)$ be the proper transform of the divisor 
$(x_3=0)\subset X_0$, explicit computation gives\footnote{Since this is a standard
computation, we again content ourselves with explaining how to perform an
equivalent computation in a more general context in Section~\ref{sec:toric}.}
\begin{equation}
  \label{eq:c2xs6}
  c_2(X^\sharp)=(2E_3+6E_2+12E_1+20L+2M+10N)N,
\end{equation}
which is visibly an even class.  Therefore $G^\sharp=0$ satisfies the
quantization condition.

We now exhibit explicit smoothings $X^\flat$ which satisfy a G-flux constraint.
In the
example under investigation, $S$ is a quintic surface and
$K_S=\mathcal{O}_S(1)$ so  we have that $K_S^2=5$.
It follows that the Euler characteristic changes by $1050=N(N-1)(N+1)K_S^2$ with $N=6$
according to eq.~\eqref{eq:Euler2}. 

Starting the transition with
$G^\sharp=0$ as above, 
then for $G^\flat$ we require $\frac12 \left(\frac{G^\flat}{2\pi}\right)^2=\frac{1050}{24}$, 
or $\left(\frac{G^\flat}{2\pi}\right)^2=\frac{175}{2}$.  We also require $G^\flat$ to satisfy 
the quantization condition~\eqref{eq:quant} that $c_2(X^\flat)-2\cdot\frac{G^\flat}{2\pi}$ is even.

We can find a suitable $G^\flat$ after
constraining $q^\flat$ to be of the form
\begin{equation}
  \label{eq:det6}
  y_0y_1=\det\left(y_2I_6+M(y)\right),
\end{equation}
where $M(y)$ is a traceless $6\times6$ matrix of linear forms in $y_3,\ldots,y_6$ and $I_6$
is the $6\times6$ identity matrix.  There are $35\times4$ moduli for the entries of $M(y)$, which must be reduced by 35 since
conjugation by an $SU(6)$ matrix does not alter $q^\flat$.  These $35\times4-35$ moduli precisely match the moduli of
the Higgs branch of an $SU(6)$ theory with 4 adjoints.  Note that
$M\equiv0$ corresponds to $q^\flat=q_0$.

Let $S(y)$ be the $6\times 5$ submatrix of $y_2I_6+M(y)$ obtained by deleting its last column.  Let
$R\subset X^\flat$ be the 4-cycle defined by $y_0=y_2=0$ and let 
$T\subset X^\flat$ be the 4-cycle defined by 
\begin{equation}
  T\,=\,\left\{y\in X^\flat \mid y_0=0\,, \ \operatorname{rank} S(y) \le 4 \right\}\ .
\end{equation}
We put
\begin{equation}
  \frac{G^\flat}{2\pi}\,=\,\frac52R-T\in H^4(X^\flat) \ ,
\end{equation}
which is of Hodge type $(2,2)$ since it is an algebraic cohomology class.  
 
We check that $G^\flat$ is primitive by computing that its image in
the cohomology of the fivefold $F$ defined by $g=0$ vanishes.  

Let $H$ be the restriction to $F$ of hyperplane class of $\mathbb{P}^{(1,5,1,1,1,1,1)}$.   Since $F$ has weight 5 and the weighted
projective space has a $\mathbb{Z}_5$ quotient, we have $\int_FH^5=5/5=1$.

Since $R$ is defined in $F$ by $q^\flat=y_0=y_2=0$, its image in $F$ is $6H^3$.
By Porteous's formula, $T$ has image $15H^3$ in $F$.  
Thus the image of the class of $(5/2)R-T$ in $F$ vanishes and
we have verified primitivity.

\smallskip
We compute $\left(\frac{G^\flat}{2\pi}\right)^2$ by computing the intersections $R^2,RT$, and $T^2$
in $X^\flat$.
Since $X^\flat$ is a $(6,5)$ complete intersection in $\mathbb{P}^{(1,5,1,1,1,1,1)}$ and
$R$ is a complete intersection of two linear forms, we have 
\begin{equation}
  R^2=\frac{6\cdot5\cdot 1^4}{5}=6,
\label{eq:r2}
\end{equation}
where the denominator of $5$ arises from the $\mathbb{Z}_5$ quotient in the weighted
projective space.

For $RT$, we can replace $R$ by the algebraically equivalent cycle $y_2=y_3=0$.  Computing $RT$ inside $F$  we get
\begin{equation}
  RT=\int_F 15H^3\cdot H^2=15.
\label{eq:rt}
\end{equation}
Finally, we compute $T^2$ as the degree of the second Chern class of the normal bundle
$N_{T,X^\flat}$ of $T$ in $X^\flat$.  First we define
\begin{equation}
\tilde{T}=\left\{(y,z)\in X^\flat\times\mathbb{P}^4\mid S(y)z=0
\right\}.
\end{equation} 
The projection $\pi:X^\flat\times\mathbb{P}^4\to X^\flat$ maps $\tilde{T}$ to $T$.  This projection fails to be an isomorphism
only over points of $T$ at which $S(y)$ has rank 3 or less.  Since the rank 3 condition is codimension~6 in $X^\flat$, we see
that $\tilde{T}\to T$ is an isomorphism.

We have
\begin{equation}\label{eq:cnt}
c(N_{T,X^\flat})=\frac{c(X^\flat)}{c(T)} \ ,
\end{equation}
where we omit restrictions to $T$ for brevity.
Similarly,
\begin{equation}
c(N_{\tilde{T},F\times\mathbb{P}^4})=\frac{c(F\times\mathbb{P}^4)}{c(\tilde{T})}=\frac{c(F)c(\mathbb{P}^4)}{c(\tilde{T})}
\end{equation}
with omitted restrictions to $\tilde{T}$.

Letting $\eta$ be the hyperplane class $\mathbb{P}^4$, we have
\begin{equation}
c(N_{\tilde{T},F\times\mathbb{P}^4})=(1+H)(1+H+\eta)^6,
\end{equation}
since  the six components of $S(y)z$ are bilinear and together with $y_0=0$ define $\tilde{T}$ as  a complete intersection. 
Since $X^\flat$ is the hypersurface in $F$ defined by $q^\flat=0$, we have
\begin{equation}
c(F)|_{X^\flat}=c(X^\flat)(1+6H).
\label{eq:cf}
\end{equation}
 Identifying $\tilde{T}$ with 
$T$ via $\pi$, we get from (\ref{eq:cnt})--(\ref{eq:cf})
\begin{equation}
  c(N_{T,X^\flat})=\frac{(1+H+\eta)^6(1+H)}{(1+\eta)^5(1+6H)}
\end{equation}
which gives
\begin{equation}
c_2(N_{T,X^\flat})=15H^2-5H\eta,
\label{eq:c2ntx}
\end{equation}
identified as a class on $\tilde{T}$.  We  can easily push (\ref{eq:c2ntx}) to
$F\times\mathbb{P}^4$ since $\tilde{T}$ 
is a complete intersection of 6 divisors in the class $H+\eta$ and the divisor $y_0=0$
of class $H$:
\begin{equation} \label{eq:c2inFP}
  c_2(N_{T,X^\flat})=(15H^2-5H\eta)H(H+\eta)^6\in H^{18}(F\times\mathbb{P}^4) 
\end{equation}
We project (\ref{eq:c2inFP}) to $F$ by extracting the coefficient of $\eta^4$, 
which is $125H^5\in H^{10}(F)$, which evaluates to 125.  So finally
\begin{equation}
T^2=125.
\label{eq:t2}
\end{equation}
Putting (\ref{eq:r2}), (\ref{eq:rt}), and (\ref{eq:t2}) together, we get
\begin{equation}
  \left(\frac{G^\flat}{2\pi}\right)^2=\frac{25}4R^2-5RT+T^2=\frac{75}2-75+125=\frac{175}2
\end{equation}
as required.

\smallskip
For the quantization condition, explicit computation gives $c_2(X^\flat)=15H^2$.
Since we only need to compute mod 2, we can replace $2\cdot\frac{G^\flat}{2\pi}=5R-2T$
by $5R=5H^2$ because $R$ is a complete intersection of two linear equations in 
$X^\flat$.  We learn that 
$c_2(X^\flat)-2\cdot\frac{G^\flat}{2\pi}$ is congruent mod 2 to 
$10H^2$, which is even. So the quantization condition is satisfied.

\medskip
Looking at (\ref{eq:det6}), the action of the Weyl group
$\mathcal{W}(SU(6))\simeq S_6$ is realized by permuting the rows and columns of $M(y)$.
As we did at the end of Section~\ref{sec:example}, we can realize this Weyl
group action as a monodromy.  We rewrite the submatrix $S(y)$ of $y_2I_6+M(y)$ as
\begin{equation}
  \label{eq:nfromm}
  S(y)=\left(
    \begin{array}{ccccc}
      1&0&0&0&0\\
0&1&0&0&0\\
0&0&1&0&0\\
0&0&0&1&0\\
0&0&0&0&1\\
0&0&0&0&0
    \end{array}
\right)\left(y_2I_6+M(y)\right).
\end{equation}
The projection matrix $P$ appearing in (\ref{eq:nfromm}) has columns $e_1,
\ldots,e_5$ chosen from the standard basis of $\mathbb{C}^6$.  Choose
a permutation $\sigma$ in the Weyl group.
We can then choose a path in the space of $6\times5$ matrices of maximal rank
starting from $P$ and ending
at the 
matrix whose columns are $e_{\sigma(1)},\ldots,e_{\sigma(5)}$ to realize the Weyl
action as a monodromy.

\section{Toric geometry and further examples}
\label{sec:toric}

We begin by reviewing the setup for investigating Calabi--Yau hypersurfaces and complete
intersections in toric varieties, specialized to Calabi--Yau fourfolds.  See 
\cite{batyrev,batbor}.

Let $N$ and $M$ be a pair of dual lattices of rank~5.  We consider a pair 
$(\Delta,\Delta^\circ)$ of 5-dimensional reflexive polytopes, with 
$\Delta\subset M_{\mathbb{R}}$ spanned by vertices in $M$, and 
$\Delta^\circ\subset N_{\mathbb{R}}$ spanned by vertices in $N$.  The origin is
the unique interior point of $\Delta\cap M$ and of $\Delta^\circ\cap N$. The polytopes
are related by
\begin{equation}
  \label{eq:polar}
  \Delta^\circ=\left\{
n\in N_{\mathbb{R}}\mid \langle m,n\rangle\ge 1\ {\rm for\ all\ }m\in\Delta
\right\}.
\end{equation}
The toric variety $\mathbb{P}_{\Delta}$ can also be described as the toric
variety associated to fan obtained by taking the cones over the faces of 
$\Delta^\circ$.  Since this toric variety is typically highly singular, we 
choose a maximal projective crepant subdivision of that fan to obtain a toric   
variety with controllable singularities.  The fan $\Sigma^\sharp$ of
this toric variety satisfies
\begin{enumerate}
\item $\Sigma^\sharp(1)=\Delta^\circ\cap N - \{0\}$
\item $X_{\Sigma^\sharp}$ is projective and simplicial
\end{enumerate}
We let $X^\sharp\subset X_{\Sigma^\sharp}$ be a general anticanonical hypersurface,
so that $X^\sharp$ is a Calabi--Yau fourfold.

\subsection{Gauge group $SU(N)$}

To achieve the situation of $A_{N-1}$ singularities, we assume:
\begin{equation}
\Delta^\circ\ \text{has a one-dimensional edge\ }\Gamma\ \text{containing\ }N-1\ \text{interior lattice points}
\end{equation}

Let $v_1$ and $v_2$ be the endpoints of $\Gamma$.  We can choose an $m_\Gamma\in M$ so that
$\langle m_\Gamma,v_1\rangle=N-1$ and $\langle
m_\Gamma,v_2\rangle=-1$.

If we remove the cones containing the interior lattice points of $\Gamma$ from the fan $\Sigma^\sharp$, we obtain a fan $\Sigma_0$.
The natural map $\pi:X_{\Sigma^\sharp}\to X_{\Sigma_0}$ blows down a divisor to a threefold
with transverse $A_{N-1}$ singularities.  After intersecting with an anticanonical
hypersurface in $X_{\Sigma^\sharp}$, we get a map $X^\sharp\to X_0$ of Calabi--Yau
fourfolds, contracting a divisor to a surface $S$ of $A_{N-1}$ singularities.

To begin to understand $S$, we consider the dual face $\Gamma^\circ\subset
\Delta$ defined by
\begin{equation}
  \label{eq:dualface}
  \Gamma^\circ=\left\{
m\in M_{\mathbb{R}}\mid 
\langle m,v_1
\rangle =-1,\ 
\langle m,v_2
\rangle=-1
\right\}.
\end{equation}
Then $\Gamma^\circ$ is a 3-dimensional polytope.  We have for the geometric genus of $S$
\begin{equation} \label{eq:pgstoric}
  p_g=|\mathrm{int}\left(\Gamma^\circ\right)\cap M| \ ,
\end{equation}
as we will check later.

We denote the primitive integral generators of the other one-dimensional
cones in $\Sigma_0$ by $v_3,\ldots,v_k$.  We also denote by $D_i\subset
X_{\Sigma_0}$ the toric divisor associated with the edge $v_i$, $1\le i\le k$.  
Similarly,
we denote by ${D}_i^\sharp\subset
X_{\Sigma^\sharp}$ the toric divisor associated with the edge $v_i$.  For 
${D}_i^\sharp$, we can have $1\le i\le k$ as above, or $v_i$ can
denote one of the $n-1$ interior lattice points of $\Gamma$.

We form a new simplicial fan ${\Sigma}^\flat$ in $(N\oplus\mathbb{Z})_{\mathbb{R}}$ with 1-dimensional cones $w_0,\ldots,w_n$ given by
by the vertices
\begin{equation}  \label{eq:newvertices}
\begin{aligned}
    w_0&=(\tfrac{v_1-v_2}N,-(N-1)) \ , \\
    w_1&=(0,1) \ , \\
    w_2&=(v_2,0) \ , \\
    w_i&=(v_i,-N\langle m_\Gamma,v_i\rangle) \ ,\ i\ge 3 \ .
\end{aligned}
\end{equation}
The six-dimensional cones $\Sigma^\flat(6)$ of $\Sigma^\flat$ can be described as follows.  Let $\sigma\in\Sigma_0(5)$ be a 5-dimensional (simplicial) cone of $\Sigma_0$.  We partition the edges $\sigma(1)$ of $\sigma$ into the set $\sigma(1)_1$ of edges spanned by $v_1$ or $v_2$, and the set
$\sigma(1)_2$ of edges spanned by $v_i$ with $i\ge3$.   We have abused notation 
slightly by labeling the edges by their primitive integral generators $v_i$.

Then we form $\Sigma^\flat(6)$ as follows: for each $\sigma\in\Sigma(5)$ we form one or more 6-dimensional simplicial cones as the span of
the vectors $w_{i-1}$ for each $v_i\in \sigma(1)_1$, together with the 
vectors $w_i$ for each $v_i\in\sigma(1)_2$, and exactly one more vector
from among $\{w_0,w_1,w_2\}$.  A distinct cone is included in $\Sigma^\flat(6)$
for each choice of this additional vector $w_0,w_1$
or $w_2$. The fan $\Sigma^\flat$ is a the fan whose cones are the faces of one of the top-dimensional
cones just described.

Using the fact that $\Sigma_0$ is a fan, it is straightforward to check that the intersection of any two cones of $\Sigma^\flat$ is
a face of each, so that $\Sigma^\flat$ is indeed a fan.  Furthermore, it is straightforward to check that $\Sigma^\flat$ is
complete since $\Sigma_0$ is.  We let $D^\flat_i\subset X_{\Sigma^\flat}$ be the
toric divisor associated with the edge $w_i$.

\medskip\noindent
{\em Example.\/} Let $\Sigma_0$ be a fan for $\mathbb{P}^{(1,1,2,2,2,2)}$.  A convenient choice is to take the complete
simplicial fan with edges spanned by the rows of
\begin{equation}
\left(\begin{array}{ccccc}
-1&-2&-2&-2&-2\\
1& 0 & 0 &0& 0\\
0& 1& 0&0&0\\
0&0&1&0&0\\
0&0&0&1&0\\
0&0&0&0&1\\
\end{array}\right)
\end{equation}
 We label the rows as $v_1,\ldots,v_6$ in order.  The edge $\Gamma$ joining $v_1$ and $v_2$  has one interior lattice point $v_0=(0,-1,-1,-1,-1)$ and we have
an $SU(2)$ example.  The fan for $\Sigma^\sharp$ is obtained from the fan
for $\Sigma$ by replacing each cone containing both $v_1$ and $v_2$  by two 
cones: one cone in which $v_1$ and $v_2$ are replaced by $v_1$ and $v_0$,
and other cone in which $v_1$ and $v_2$ are replaced by $v_0$ and $v_2$.

Choosing $m_\Gamma=(-1,0,0,0,0)\in M$ we have 
\begin{equation}
\langle m_\Gamma,v_1\rangle=1,\ \langle m_\Gamma,v_2\rangle=-1,\ \langle m_\Gamma,v_i\rangle=0\ {\rm for\ }
i\ge 3.
\end{equation}
Then (\ref{eq:newvertices}) gives the edges of the fan $\Sigma^\flat$ as the rows of
\begin{equation}
\left(\begin{array}{cccccc}
-1&-1&-1&-1&-1&-1\\
0&0&0&0&0&1\\
1&0&0&0&0&0\\
0&1&0&0&0&0\\
0&0&1&0&0&0\\
0&0&0&1&0&0\\
0&0&0&0&1&0
\end{array}\right)
\end{equation}
with the labeling $w_0,\ldots,w_6$.  The cones of $\Sigma^\flat$ are immediately seen to consist of all cones spanned by
any proper subset of $\{w_0,\ldots,w_6\}$.  We therefore obtain the fan for $\mathbb{P}^{6}$, the space that we embedded
$\mathbb{P}^{(1,1,2,2,2,2)}$ into
in Section~\ref{subsubsec:su2geom}.

\medskip\noindent
{\em Example.\/} Let $\Sigma_0$ be a fan for $\mathbb{P}^{(1,5,6,6,6,6)}$.  A convenient choice is to take the complete
simplicial fan with edges spanned by the rows of
\begin{equation}
\left(\begin{array}{ccccc}
-5&-6&-6&-6&-6\\
1& 0 & 0 &0& 0\\
0& 1& 0&0&0\\
0&0&1&0&0\\
0&0&0&1&0\\
0&0&0&0&1\\
\end{array}\right)
\end{equation}
 We label the rows as $v_1,\ldots,v_6$ in order.  The edge $\Gamma$ joining $v_1$ and $v_2$  has 5 interior lattice points and we have an $SU(6)$ example.
The interior lattice points are
\begin{equation}
  \label{eq:intlp}
\begin{aligned}
  &v_0=  (0,-1,-1,-1,-1)\, , \ v_{-1}= (-1,-2,-2,-2,-2)\, , \ v_{-2}=(-2,-3,-3,-3,-3) \, ,\\
  &v_{-3}= (-3,-4,-4,-4,-4)\,,\  v_{-4}=(-4,-5,-5,-5,-5,-5)\, .
\end{aligned}
\end{equation}

Choosing $m_\Gamma=(-1,0,0,0,0)\in M$ we have 
\begin{equation}
\langle m_\Gamma,v_1\rangle=5,\ \langle m_\Gamma,v_2\rangle=-1,\ \langle m_\Gamma,v_i\rangle=0\ {\rm for\ }
i\ge 3.
\end{equation}
Then (\ref{eq:newvertices}) gives the edges of the fan $\Sigma^\flat$ as the rows of
\begin{equation}
\left(\begin{array}{cccccc}
-1&-1&-1&-1&-1&-5\\
0&0&0&0&0&1\\
1&0&0&0&0&0\\
0&1&0&0&0&0\\
0&0&1&0&0&0\\
0&0&0&1&0&0\\
0&0&0&0&1&0
\end{array}\right)
\end{equation}
with the labeling $w_0,\ldots,w_6$.  The cones of $\Sigma^\flat$ are immediately seen to consist of all cones spanned by
any proper subset of $\{w_0,\ldots,w_6\}$.  We therefore obtain the fan for $\mathbb{P}^{(1,1,1,1,1,1,5)}$, the space that we embedded
$\mathbb{P}^{(1,5,6,6,6,6)}$ into
in Section~\ref{subsubsec:su6geom}.\footnote{The general toric procedure 
requires us to add more edges from additional points of $\Delta^\circ\cap N$.  
However, for simplicity we can safely exclude them from discussion since 
the weighted hypersurface $f$ considered in Section~\ref{subsubsec:su6geom}
does not contain the fixed point of the $\mathbb{P}^5$ action.}

Consider the map $\iota:X_{\Sigma_0}\to X_{{\Sigma}^\flat}$ given by
  \begin{equation}
(y_0,y_1,y_2,y_3,\ldots)=(x_1^N,x_2^N,x_1x_2,x_3\ldots),
\label{eq:embed}
  \end{equation}
where $(y_0,\ldots,y_k)$ are the homogeneous coordinates of $X_{\Sigma^\flat}$.
We check that the map is well-defined. If $g=(t_j)\in G(\Sigma_0)$, then
$\prod t_j^{\langle m,v_j\rangle}=1$ for all $m\in M$.  Then $g\cdot x$ maps to
\begin{equation}
  (t_1^Nx_1^N,t_2^Nx_2^N,t_1t_2x_1x_2,t_3x_3\ldots),
\end{equation}
which we have to show is equivalent to 
$(x_1^N,x_2^N,x_1x_2,\ldots)$ up to an element of
$G({\Sigma}^\flat)$.  In other words, we have to check that for all $(m,n)\in M\oplus
\mathbb{Z}$ we have
\begin{equation}
  (t_1^N)^{\langle (m,n),w_0\rangle}
(t_2^N)^{\langle (m,n),w_1\rangle}
  (t_1t_2)^{\langle (m,n),w_2\rangle}\prod_{j=3}^k
  (t_j)^{\langle (m,n),w_j\rangle}=1.
\label{eq:checkgf}
\end{equation}
But the left hand side of (\ref{eq:checkgf}) simplifies to
\begin{equation}
  \label{eq:checksimp}
  t_1^{\langle m,v_1\rangle-N(N-1)n}t_2^{\langle m,v_2\rangle+Nn}\prod_{j=3}^k
  (t_j)^{\langle (m-Nnm_\Gamma,v_j\rangle}=\prod_{j=1}^k
  (t_j)^{\langle (m-Nnm_\Gamma,v_j\rangle},
\end{equation}
which is 1 because $m-Nnm_\Gamma\in M$ and $g\in G(\Gamma)$.

It is straightforward to check that the image of $\iota$ in coordinates
lands in $\mathbb{C}^{k+1}-Z(\Sigma^\flat)$ and is an embedding after modding
out by $G(\Sigma_0)$ and $G(\Sigma^\flat)$.

\smallskip\noindent
{\em Example.} For $\mathbb{P}^{(1,1,2,2,2,2)}$ the embedding into $\mathbb{P}^{6}$ is
\begin{equation}
(y_1,\ldots,y_6)\mapsto (y_1^2,y_2^2,y_1y_2,y_3,\ldots,y_6)
\end{equation}
in complete agreement with Section~\ref{subsubsec:su2geom}.

\smallskip\noindent
{\em Example.} For $\mathbb{P}^{(1,5,6,6,6,6)}$ the embedding into $\mathbb{P}^{(1,5,1,1,1,1,1)}$ is
\begin{equation}
(y_1,\ldots,y_6)\mapsto (y_1^6,y_2^6,y_1y_2,y_3,\ldots,y_6)
\end{equation}
in complete agreement with Section~\ref{subsubsec:su6geom}.

Clearly, $\iota(X_{\Sigma_0})$ is contained in the hypersurface $
q_0(y)=y_0y_1-y_2^N=0$.   The linear equivalence $D^\flat_0+D^\flat_1\sim ND^\flat_2$ is realized by $(Nm_\Gamma,1)\in M\oplus\mathbb{Z}$.

 Since $\iota^*(\mathcal{O}_{X_{\Sigma^\flat}}(D^\flat_2))\simeq
\mathcal{O}_{X_{\Sigma_0}}(D_0+D_1)$ and  
$\iota^*(\mathcal{O}_{X_{\Sigma^\flat}}(D^\flat_j))\simeq
\mathcal{O}_{X_{\Sigma_0}}(D_j)$ for $j\ge 3$, we see that 
\begin{equation}
\iota^*(\mathcal{O}_{X_{\Sigma^\flat}}(\sum_{j=2}^kD^\flat_j))\simeq
\mathcal{O}_{X_{\Sigma_0}}(\sum_{j=1}^kD_j)=\mathcal{O}_{X_{\Sigma_0}}(-K_{X_\Sigma}),
\end{equation}
and it is easy to see that $X_0$ pulls back from a section of
$\mathcal{O}_{X_{\Sigma^\flat}}(\sum_{j=2}^kD^\flat_j)=:\mathcal{O}_{X_{\Sigma^\flat}}(D')$ which we denote by
$f(y)$.  Thus
$X_0$ is identified with a complete intersection of $q_0(y)$
and $f(y)$ in $X_{\Sigma^\flat}$.  The singular locus $S$ is defined by $y_0=y_1=y_2=f(y)=0$.  Adjunction again
says that the canonical bundle of $S$ is the restriction of $D^\flat_2$.

We can describe this complete intersection using a nef partition
\cite{batbor} if desired, partitioning the edges $\rho_j$ spanned by the $w_j$ 
into two sets:
\begin{equation}
\{\rho_0,\rho_1\},\{\rho_2, \rho_3,\ldots,\rho_k\}
\end{equation}
and we are led to view $q_0(y)$ as a section of 
$\mathcal{O}_{X_{\Sigma^\flat}}(D_0+D_1)$.

We can identify a basis of sections of $\mathcal{O}_{X_{\Sigma^\flat}}(D^\flat_2)$
with monomials $\chi^{(m,n)}$, $(m,n)\in M\oplus\mathbb{Z}$, satisfying
\begin{equation}
  \begin{array}{ccccl}
  \langle (m,n),w_0\rangle&=&\langle m,\frac{v_1-v_2}2\rangle-n&\ge&0\\
  \langle (m,n),w_1\rangle&=&n&\ge&0\\
  \langle (m,n),w_2\rangle&=&\langle m,v_2\rangle&\ge&-1\\
  \langle (m,n),w_j\rangle&=&\langle m,v_j\rangle&\ge&0\ (j\ge3)\\
  \end{array}
\label{eq:d2ncond}
\end{equation}
via the correspondence
\begin{equation}
  \chi^{(m,n)}\leftrightarrow z^{(m,n)}:=y_2\prod_{j=0}^ky_j^{\langle (m,n),w_k\rangle}
\end{equation}

If $m\in \mathrm{int}(\Gamma^\circ)\cap M$, (\ref{eq:d2ncond}) is satisfied for $(m,0)\in M\oplus\mathbb{Z}$, 
since $m\in \mathrm{int}(\Gamma^\circ)\cap M$ is equivalent to
\begin{equation}
\langle m,v_1\rangle= -1,\ \langle m,v_2\rangle= -1,\ 
\langle m,v_j\rangle\ge 0\ (j\ge 3).
\end{equation}
Thus $y_0y_1,y_2^N$, and $y_2^r\prod_{j=1}^{N-r}z^{(m_{i_j},0)}$ ($m_{i_j}\in \mathrm{int}(\Gamma^\circ)\cap M$)
are identified with sections of $\mathcal{O}_{X_{\Sigma^\flat}}(ND^\flat_2)$.  By the now-familiar argument, we will only need
to consider $r\le N-2$.

We change notation and rewrite the $p_g$ sections
$z^{(m_i,0)}$, $m_i\in\mathrm{int}(\Gamma^\circ)\cap M$ as $z_1,\ldots,z_{p_g}$.  Then
\begin{equation}
  \label{eq:qdef}
  q^\flat(y) \,=\, y_0y_1+y_2^N+\sum_{r=0}^{N-2}y_2^r\sum_{\stackrel{J\subset\{1,\ldots,p_g\}}{|J|=N-r}}
 a_{j_1\ldots j_{N-r}}
\prod z_{j_i}
\end{equation}
is a deformation of $q_0(y)$.

\medskip\noindent
Suppose that $c_2(X^\sharp)$ is the restriction of an even
toric class in $H^4(X_{\Sigma^\sharp},\mathbb{Z})$, which implies that $G^\sharp=0$ satisfies
quantization.  Then we 
will exhibit explicit smoothings $X^\flat$ which satisfy a G-flux constraint.  

\smallskip
The strengthened hypothesis on the evenness of $c_2(X^\sharp)$ 
is needed so that the exhibited $G^\flat$ satisfies the 
quantization condition.  The examples from Section~\ref{sec:wpex} both satisfy
this hypothesis.  The tadpole condition is always satisfied, as we will see.

\medskip
We can find a suitable $G^\flat$ after
constraining $q^\flat$ to be of the form
\begin{equation}
  \label{eq:detn}
  y_0y_1=\det\left(y_2I_N+M(y)\right),
\end{equation}
where $M(y)$ is a traceless $N\times N$ matrix of linear forms in $y_3,\ldots,y_{p_g+2}$ and $I_N$
is the $N\times N$ identity matrix.  There are $p_g(N^2-1)$ moduli for the entries of $M(y)$, which must be reduced by $N^2-1$ since
conjugation by an $SU(N)$ matrix does not alter $q^\flat$.  These $(p_g-1)(N^2-1)$ moduli precisely match the moduli of
the Higgs branch of an $SU(N)$ theory with $p_g$ adjoints.  Note that
$M\equiv0$ corresponds to $q^\flat=q_0$.

Let $S(y)$ be the $N\times (N-1)$ submatrix of $y_2I_N+M$ obtained by deleting its last column.  Let
$R\subset X^\flat$ be the 4-cycle defined by $y_0=y_2=0$ and let 
$T\subset X^\flat$ be the 4-cycle defined by 
\begin{equation}
  T=\left\{y\in X^\flat\mid y_0=0\,, \  \operatorname{rank} S(y) \le N-2
\right\}.
\end{equation}
We put
\begin{equation}
  \frac{G^\flat}{2\pi}\,=\,\frac{N-1}2R-T\in H^4(X^\flat) \ ,
\end{equation}
which is of type $(2,2)$ since it is an algebraic cohomology class.  
 
We check that $G^\flat$ is primitive by computing that its image in
the cohomology of the fivefold $F$ defined by $g=0$ vanishes.

Since $R$ is defined in $F$ by $q^\flat=y_0=y_2=0$, its class in $F$ is $ND^\flat_0(D^\flat_2)^2$.
By Porteous's formula, $T$ has class $\frac{N(N-1)}2D^\flat_0(D^\flat_2)^2$.  Thus the class of $\frac{N-1}2R-T$ vanishes in $F$ and
we have verified primitivity.

We compute $\left(\frac{G^\flat}{2\pi}\right)^2$ by computing $R^2,RT$, and $T^2$.
Since $X^\flat$ is a complete intersection of divisors in the classes $ND^\flat_2$ and $D'$,
$R$ is a complete intersection of $y_0$ and $y_2$, we have
\begin{equation}
  R^2=N(D^\flat_0)^2(D^\flat_2)^3D'\in H^{12}(X_{\Sigma^\flat}).
\end{equation}

Computing $RT$ inside $F$  we get
\begin{equation}
  RT=\frac{N(N-1)}2(D^\flat_0)^2(D^\flat_2)^3D'\in H^{10}(F).
\end{equation}
Finally, we compute $T^2$ as the degree of the second Chern class of the normal bundle
$N_{T,X^\flat}$ of $T$ in $X^\flat$.  First we define
\begin{equation}
\tilde{T}=\left\{(y,z)\in X^\flat\times\mathbb{P}^{N-2}\mid S(y)z=0
\right\}.
\end{equation} 
The projection $\pi:X^\flat\times\mathbb{P}^{N-2}\to X^\flat$ maps $\tilde{T}$ to $T$.  This projection fails to be an isomorphism
only over points of $T$ at which $S(y)$ has rank $N-3$ or less.  Since the rank 3 condition is codimension~6 in $X^\flat$, we see
that $\tilde{T}\to T$ is an isomorphism.

We have
\begin{equation}
c(N_{T,X^\flat})=\frac{c(X^\flat)}{c(T)} \ ,
\end{equation}
where we omit restrictions to $T$ for brevity.
Also,
\begin{equation}
c(N_{\tilde{T},F\times\mathbb{P}^{N-2}})=\frac{c(F\times\mathbb{P}^{N-2})}{c(\tilde{T})}=\frac{c(F)c(\mathbb{P}^{N-2})}{c(\tilde{T})}=\frac{c(X^\flat)(1+ND^\flat_2)c(\mathbb{P}^{N-2})}{c(\tilde{T})}.
\end{equation}
We get
\begin{equation}
c(N_{\tilde{T},F\times\mathbb{P}^{N-2}})=(1+D^\flat_0)(1+D^\flat_2+\eta)^{N},
\end{equation}
since  the $N$ components of $S(y)z$ are bilinear in $\mathbb{P}^{N-2}$ and sections of $\mathcal{O}(D^\flat_2)$, which define $\tilde{T}$ as  a complete intersection together with $y_0$.  Identifying $\tilde{T}$ with 
$T$ via $\pi$, we get
\begin{equation}
  c(N_{T,X^\flat})=\frac{(1+D^\flat_2+\eta)^N(1+D^\flat_0)}{(1+ND^\flat_2)(1+\eta)^{N-1}}
\end{equation}
which gives
\begin{equation}
c_2(N_{T,X^\flat})=(D^\flat_0-ND^\flat_1)\eta+\frac{N(N-1)}2(D^\flat_2)^2=-D^\flat_1\eta+\frac{N(N-1)}2(D^\flat_2)^2,
\end{equation}
where we have used $D^\flat_0+D^\flat_1\sim N D^\flat_2$.
 Computing the intersection on $F\times\mathbb{P}^{N-2}$, this is just 
\begin{equation}
(-D^\flat_1\eta+\frac{N(N-1)}2(D^\flat_2)^2) D^\flat_0(D^\flat_2+\eta)^N\in H^{2N+6}(F\times\mathbb{P}^{N-2}).
\end{equation}
We project down to $T$ by extracting the coefficient of $\eta^{N-2}$, which is
\begin{equation}
{N\choose2}^2D^\flat_0(D^\flat_2)^4-{N\choose3}D^\flat_0D^\flat_1(D^\flat_2)^3\in H^{10}(F).
\end{equation}
Expressing this as a class on the toric variety finally gives
\begin{equation}
T^2=\left({N\choose2}^2D^\flat_0(D^\flat_2)^4-{N\choose3}D^\flat_0D^\flat_1(D^\flat_2)^3\right)D'\in H^{12}
(X_{\Sigma^\flat}).
\end{equation}
Finally
\begin{equation}
  \left(\frac{G^\flat}{2\pi}\right)^2\,=\,\frac{(N-1)^2}4R^2-(N-1)RT+T^2 \ ,
\end{equation}
which evaluates on $X_{\Sigma^\flat}$ to
\begin{equation}
\frac{N(N-1)^2}4(D^\flat_0)^2(D^\flat_2)^3D'-\frac{N(N-1)^2}2(D^\flat_0)^2(D^\flat_2)^3D'+{N\choose2}^2D^\flat_0(D^\flat_2)^4D'-{N\choose3}D^\flat_0D^\flat_1(D^\flat_2)^3D',
\end{equation}
which simplifies to 
\begin{equation}
\left(\frac{G^\flat}{2\pi}\right)^2=\frac{(N+1)N(N-1)}{12}D^\flat_0D^\flat_1(D^\flat_2)^3D'\in H^{12}(X_{\Sigma^\flat}),
\end{equation}
where we have again used $D^\flat_0+D^\flat_1\sim ND^\flat_2$.

  Since $S$ is a complete intersection of $y_0,y_1,y_2$, and $g$,
the class of $S$ is $D^\flat_0D^\flat_1D^\flat_2D'$.  By adjunction, we find
\begin{equation}
K_S=-\sum_{i=0}^k D^\flat_i+D^\flat_0+D^\flat_1+D^\flat_2+D'=D^\flat_2.
\end{equation}
Thus we get
\begin{equation}
\frac12\int_{X^\flat} \frac{G^\flat}{2\pi}\wedge\frac{G^\flat}{2\pi}\,=\,\frac{(N+1)N(N-1)}{24} K_S^2=\frac{\chi(X^\flat)-\chi(X^\sharp)}{24} \ ,
\end{equation}
as required according to eq.~\eqref{eq:fluxcont}.

\medskip
It remains to check the quantization condition. 
We first recall the computation of the Chern classes of a toric variety.  Let $X_{\Sigma}$
be a smooth projective toric variety of dimension $n$ with $\ell$ edges in the fan $\Sigma$.
Let $D_1,\ldots,D_\ell\subset X_{\Sigma}$ be the corresponding toric divisors.  Then we have
a short exact sequence\cite{fulton}
\begin{equation}
  \label{eq:torict}
  0\to \mathcal{O}_{X_\Sigma}^{\ell-n}\to \bigoplus_{i=1}^\ell\mathcal{O}_{X_\Sigma}(D_i)\to 
T_{X_\Sigma}\to 0,
\end{equation}
which gives
\begin{equation}
  \label{eq:ctxs}
  c(X_\Sigma)=\prod_{i=1}^\ell(1+D_i).
\end{equation}
So if $X\subset X_\Sigma$ is an anticanonical hypersurface we have
\begin{equation}
  \label{eq:ctx}
  c(X)=\frac{\prod_{i=1}^\ell(1+D_i)}{1+\sum_{i=1}^\ell D_i},
\end{equation}
which gives
\begin{equation}
  \label{eq:c2x}
  c_2(X)=\sum_{i<j}D_iD_j.
\end{equation}

In particular, for $X^\sharp\subset X_{\Sigma^\sharp}$ we have
\begin{equation}
  \label{eq:c2xs}
  c_2(X^\sharp)=\sum_{i<j}D^\sharp_iD^\sharp_j.
\end{equation}
Similarly, for $X^\flat$ we get
\begin{equation}
  \label{eq:cxf}
  c(X^\flat)=\frac{\prod_{i=0}^k(1+D^\flat_i)}{(1+D^\flat_0+D^\flat_1)(1+\sum_{i=2}^k
D^\flat_i)},
\end{equation}
which gives
\begin{equation}
  \label{eq:c2xf}
  c_2(X^\flat)=D^\flat_0D^\flat_1+\sum_{2\le i<j\le k}D^\flat_iD^\flat_j.
\end{equation}

We now let $f=\iota\circ\pi:X_{\Sigma^\sharp}\to X_{\Sigma^\flat}$ be the composition.
Since (\ref{eq:c2xs}) and (\ref{eq:c2xf}) show that $c_2(X^\sharp)$ is the restriction of
a cohomology class on $X_{\Sigma^\sharp}$ and 
$c_2(X^\flat)$ is the restriction of
a cohomology class on $X_{{\Sigma}^\flat}$, we are able to compare $c_2$ on both sides of
the transition using $f^*$.

We continue our labeling conventions, so that the vertices in $\Gamma$ are labeled, in order
\begin{equation}
  \label{eq:labeling}
  v_1,\ v_0,\ v_{-1},\ldots,v_{2-N},\ v_2. 
\end{equation}

We compute
\begin{equation}
  \label{eq:pullbacks}
  \begin{aligned}
    f^*(D_0^\flat)&=ND_1^\sharp+(N-1)D_0^\sharp+\ldots+D_{2-N}^\sharp \ ,\\
f^*(D_1^\flat)&=D_0^\sharp+\ldots+(N-1)D_{2-N}^\sharp+ND_2^\sharp\ , \\
f^*(D_2^\flat)&=D_1^\sharp+D_0^\sharp+\ldots+D_{2-N}^\sharp+D_2^\sharp\ , \\
f^*(D_i^\flat)&=D_i^\sharp\ ,\quad i\ge3 \ . 
  \end{aligned}
\end{equation}
As a check, note that $f^*(D_0^\flat+D_1^\flat-ND_2^\flat)=0$, as it had to be  owing
to the linear equivalence $D_0^\flat+D_1^\flat\sim ND_2^\flat$.  

Since (\ref{eq:pullbacks}) is a standard toric calculation, we content 
ourselves with just a few words of explanation.  We have 
$f^*=\pi^*\circ\iota^*$, and $\iota^*$ is calculated below in (\ref{eq:pullbackdown}).  Since $\pi$ is a blowdown, all that $\pi^*$ can do is introduce
the exceptional divisors with multiplicities.   It then follows from the
fact that the vertices (\ref{eq:labeling}) are on the edge $\Gamma$, in order,
that for a pullback the coefficients of $D_1^\sharp,D_0^\sharp,\ldots,
D_{2-N}^\sharp,D_2^\sharp$ must be in arithmetic progression.  Since the
coefficients of $D_1^\sharp$ and $D_2^\sharp$ are fixed by (\ref{eq:pullbackdown}),
these observations are enough to completely determine (\ref{eq:pullbacks}).

For the quantization condition, we only need to compute mod 2.  
Since $T$ is an integral class, we can replace  $\frac{N-1}2R-T$ with $\frac{N-1}2R$ in checking
quantization.  We see that quantization follows immediately from two claims:
\begin{itemize}
\item $f^*(c_2(X^\flat)-(N-1)R)\equiv c_2(X^\sharp)\ (\mathrm{mod}\ 2)$ 
\item $f^*:H^*(X_{{\Sigma}^\flat},\mathbb{Z}_2)\to H^*(X_{\Sigma^\sharp},\mathbb{Z}_2)$ is 
injective
\end{itemize}

The first claim is checked by direct calculation, which can be separated into the cases 
where $N$ is odd and $N$ is even.

If  $N$ is odd, then (\ref{eq:pullbacks}) simplifies to
\begin{equation}
  \label{eq:pullbacksodd}
  \begin{aligned}
    f^*(D_0^\flat)&=D_1^\sharp+D_{-1}^\sharp+\ldots+D_{4-N}^\sharp+D_{2-N}^\sharp\ ,\\
f^*(D_1^\flat)&=D_0^\sharp+D_{-2}^\sharp+\ldots+D_{3-N}^\sharp+D_2^\sharp\ ,\\
f^*(D_2^\flat)&=D_1^\sharp+D_0^\sharp+\ldots+D_{2-N}^\sharp+D_2^\sharp\ ,\\
f^*(D_i^\flat)&=D_i^\sharp\ ,\quad i\ge3\ . 
  \end{aligned}
\end{equation}
Also, since $N-1$ is even, we only have to show that $f^*(c_2(X^\flat))\equiv
c_2(X^\sharp)\ (\mathrm{mod}\ 2)$.  This follows immediately from 
(\ref{eq:c2xs}), (\ref{eq:c2xf}), (\ref{eq:pullbacksodd}), and the Stanley-Reisner
vanishings:
\begin{equation}
  \label{eq:SRodd}
  D_i\cdot D_j=0\ {\rm for\ } 2-N\le i<j\le 2\ {\rm unless\ }v_i,\ v_j\ {\rm are\ 
adjacent\ in\ the\ ordering\ (\ref{eq:labeling})} 
\end{equation}

Similarly, if $N$ is even we have
\begin{equation}
  \label{eq:pullbackseven}
  \begin{aligned}
    f^*(D_0^\flat)&=D_0^\sharp+D_{-2}^\sharp+\ldots+D_{2-N}^\sharp\ ,\\
f^*(D_1^\flat)&=D_0^\sharp+D_{-2}^\sharp+\ldots+D_{2-N}^\sharp\ ,\\
f^*(D_2^\flat)&=D_1^\sharp+D_0^\sharp+\ldots+D_{2-N}^\sharp+D_2^\sharp\ ,\\
f^*(D_i^\flat)&=D_i^\sharp\ ,\quad i\ge3 \ . 
  \end{aligned}
\end{equation}
Then $(N-1)R\equiv R\ (\mathrm{mod}\ 2)$.  Since $R=D_0^\flat D_2^\flat$, the claim
follows from $f^*(c_2(X^\flat)+D_0^\flat D_2^\flat)\equiv c_2(X^\sharp)\ (\mathrm{mod}\ 2)$,
which is again checked by direct calculation as above. 

\smallskip
The injectivity of $f^*$ can be broken down into the injectivity of $\iota^*$ and $\pi^*$
separately.  The injectivity of $\pi^*$ follows since $\pi$ is a blowup.  The injectivity
of $\iota^*$ follows from the simpler computation analogous to (\ref{eq:pullbacks})
\begin{equation}
  \label{eq:pullbackdown}
  \begin{aligned}
    \iota^*(D_0^\flat)&=ND_1\ ,\\
\iota^*(D_1^\flat)&=ND_2\ ,\\
\iota^*(D_2^\flat)&=D_1+D_2\ ,\\
\iota^*(D_i^\flat)&=D_i\ ,\quad i\ge3 \ ,
  \end{aligned}
\end{equation}
which follows immediately from (\ref{eq:embed}).
The form of the fan $\Sigma^\sharp$ shows that all linear equivalences  and Stanley-Reisner
relations among the $D_i$ pull back from
from corresponding relations in the $D_i^\flat$.  So we only have to look at
(\ref{eq:pullbackdown}) as a linear transformation to deduce that the kernel of
$\iota^*$ is generated by $D_0^\flat+D_1^\flat-ND_2^\flat$, which is zero.

\section{Conclusions} \label{sec:conc}
In this work we studied the three-dimensional $\mathcal{N}=2$ low energy theory of M-theory on Calabi--Yau fourfolds $X_0$ with a smooth surface $S$ of $A_{N-1}$ singularities. We found that --- due to massless M2-brane degrees of freedom from the $A_{N-1}$ singularity at codimension two --- the three-dimensional effective theory resulted in a $\mathcal{N}=2$ $SU(N)$ gauge theory with adjoint matter multiplets at low energies. Alternatively, we obtained the same gauge theory from a twisted dimensional reduction of the seven-dimensional $\mathcal{N}=1$ $SU(N)$ gauge theory on the surface $S$.

From the twisted dimensional reduction, we derived for the three-dimensional $\mathcal{N}=2$ $SU(N)$ gauge theory its matter spectrum, consisting of adjoint-valued $\mathcal{N}=2$ chiral multiplets. Furthermore, we established that a variant of the Vafa--Witten equations \cite{Vafa:1994tf} governed the supersymmetric ground states of the low energy theory.  These equations allowed us to determine the moduli spaces of the Higgs and Coulomb branches of the gauge theory, where we in particular focus on the twisted dimensional reduction on $S$ with a trivial $SU(N)$ principal bundle.

From the results of the performed gauge theory analysis, we predicted geometric properties of the M-theory compactification on the singular Calabi--Yau fourfold $X_0$. First of all, we matched Coulomb and Higgs branches of the gauge theory with the crepant resolution to the (smooth) Calabi--Yau fourfolds~$X^\sharp$ and with the deformation to the (smooth) Calabi--Yau fourfold~$X^\flat$, respectively. That is to say, a transition from the Coulomb to the Higgs branch in the gauge theory corresponded to an extremal transition between the resolved Calabi--Yau fourfold $X^\sharp$ and the deformed Calabi--Yau fourfold $X^\flat$ in M-theory. Furthermore, we argued that in order to arrive at the anticipated $SU(N)$ gauge theory branches --- arising from a trivial $SU(N)$ principal bundle over $S$ --- the Coulomb--Higgs phase transition starting from a Calabi--Yau fourfold~$X^\sharp$ with no background flux ends at a Calabi--Yau fourfold $X^\flat$ with non-trivial background four-form flux~$G^\flat$. 

The proposed flux $G^\flat$ was required for consistency reasons so as to match the tadpole cancellation condition --- due to the change of Euler characteristic along the extremal transition \cite{Intriligator:2012ue} --- and to fulfill the flux quantization condition of M-theory \cite{Witten:1996md}. But maybe even more importantly, the correct choice of the flux $G^\flat$ was essential to be in accord with the moduli space of the Higgs branch of the $SU(N)$ gauge theory. Namely, we showed that the background flux $G^\flat$ was primitive and generated a non-trivial M-theory superpotential. The flux $G^\flat$ was of Hodge type $(2,2)$ along the flat directions of the flux-induced superpotential, which in turn comprised the unobstructed complex structure moduli deformations associated to the Higgs branch of the described $SU(N)$ gauge theory. Furthermore, we observed that as we moved about the M-theory moduli space in the Calabi--Yau phase associated to the gauge theory Higgs branch, the flux~$G^\flat$ exhibited non-trivial monodromy behavior given in terms of the described Weyl group action $\mathcal{W}(SU(N))$ on the flux $G^\flat$.

In order to demonstrate our general arguments --- inspired by refs.~\cite{MR2092771,MR2169828} --- using the framework of toric geometry we explicitly gave examples for extremal transitions between the Calabi--Yau fourfolds $X^\sharp$ and $X^\flat$. Namely, starting from a five-dimensional toric varieties with $A_{N-1}$ singularities in codimension two, we realized the Calabi--Yau fourfold $X^\sharp$ as a hypersurface in the resolved toric variety, whereas we constructed the deformed Calabi--Yau fourfold~$X^\flat$ as a complete intersection in a six-dimensional toric variety. With the toric computational tools at hand, these examples allowed us to explicitly verify the general predictions concerning the interplay between the gauge theory moduli spaces and the M-theory background fluxes.

In this work we mainly focused on a particular gauge theory scenario arising from the twisted dimensional reduction of a trivial $SU(N)$ principal bundle over surface $S$.  Firstly, extending the analysis to non-trivial $SU(N)$ principal bundles over the surface $S$ would correspond to extremal transitions in M-theory with non-trivial background fluxes on both Calabi--Yau fourfolds $X^\sharp$ and $X^\flat$ --- in analogy to the findings for M-theory four-form fluxes associated to phases of three-dimensional $\mathcal{N}=2$ Abelian gauge theories~\cite{Intriligator:2012ue}. Secondly, it would be interesting to extend the analysis to general ADE  or even non-simply laced gauge groups. Note also that since the obtained results depended only on the local geometry in the vicinity of a codimension two singularity in the singular Calabi--Yau fourfold~$X_0$, the gauge theory branches are already captured in M-theory in terms of extremal transitions among suitable local Calabi--Yau fourfolds. Thus the relevant local Calabi--Yau fourfolds deserve further study as well. We plan to return to these issues in the future \cite{WProg}.

\section*{Acknowledgments}
We would like to thank
Lara Anderson, Paul Aspinwall,
Sergei Gukov,
Jonathan Heckman,
Ken Intriligator, 
and Peter Mayr
for discussions and correspondences. 
S.K, D.R.M., and M.R.P. thank the Aspen Center for Physics, which is
supported by NSF grant PHY-1066293,  for hospitality
during the initial stages of this work.
The work of S.K. is supported by NSF grants DMS-12-01089 and DMS-15-02170,
the work of D.R.M. is supported by NSF grant PHY-1307513, and 
the work of M.R.P. is supported by NSF grant PHY-1521053.


\ifx\undefined\bysame
\newcommand{\bysame}{\leavevmode\hbox to3em{\hrulefill}\,}
\fi

\end{document}